\documentclass[vecphys]{svmult}
\usepackage{amssymb}
\usepackage{makeidx}
\usepackage{graphicx}
\usepackage{multicol}

\authorrunning{Efetov, Garifullin, Volkov, Westerholt}
\titlerunning{S/F proximity effect}
\makeindex
\begin{document}

\title*{Proximity Effects in Ferromagnet/Superconductor Heterostructures}
\author{K. B. Efetov\inst{1,2}, I. A. Garifullin\inst{3}, A. F. Volkov%
\inst{1,4} \and K. Westerholt\inst{5}}

\institute{Institut f\"ur Theoretische Physik III,
Ruhr-Universit\"at Bochum D-44780 Bochum, Germany
\texttt{efetov@tp3.rub.de} \and L. D. Landau Institute for
Theoretical Physics RAS, 119334 Moscow, Russia \and Zavoisky
Physical-Technical Institute RAS, 420029 Kazan, Russia
\texttt{ilgiz\_garifullin@yahoo.com} \and Institute for
Radioengineering and Electronics RAS, 125009 Moscow, Russia
\texttt{volkov@tp3.rub.de} \and Institut f\"ur
Experimentalphysik/Festk\"orperphysik, Ruhr-Universit\"at Bochum
D-44780 Bochum, Germany \texttt{kurt.westerholt@rub.de}}

\maketitle \textbf{Abstract}

We review the present status of the experimental and theoretical research on
the proximity effect in heterostructures composed of superconducting (S) and
ferromagnetic (F) thin films. First, we discuss traditional effects
originating from the oscillatory behavior of the superconducting pair wave
function in the F-layer. Then, we concentrate on recent theoretical
predictions for S/F layer systems. These are a) generation of odd triplet
superconductivity in the F-layer and b) ferromagnetism induced in the
S-layer below the superconducting transition temperature $T_{c}$ (inverse
proximity effect). The second part of the review is devoted to discussion of
experiments relevant to the theoretical predictions of the first part. In
particular, we present results of measurements of the critical temperature $%
T_{c}$ as a function of the thickness of F-layers and we review
experiments indicating existence of the odd triplet
superconductivity, cryptoferromagnetism and inverse proximity
effect.

\newpage

\section{Introduction}

If a superconducting layer S is brought into contact with a non
superconducting metallic layer N, the superconducting critical temperature $%
T_{c}$ of S decreases with increasing the\textbf{\ }thickness of the N-layer
and the superconducting condensate penetrates into the N-layer over a long
distance. This phenomenon, the conventional \textit{proximity effect}, has
been studied for many years starting from the beginning of the 1960's (see
reviews \cite{deGennes,Deutscher}).

Although attractive electron-electron interaction may be absent in the
N-layer, the condensate wave function (or the Cooper pair wave function) $%
f(t-t^{\prime })$ penetrates into N over a distance $\xi _{N}$, much
exceeding\textbf{\ }the interatomic spacing. In a diffusive metal, i.e. in a
metal with a high impurity concentration ($\tau T/\hbar <<1$, where $\tau $
is the momentum relaxation time), the length $\xi _{N}$, called the
correlation length, is given by $\xi _{N}=\sqrt{D/2\pi T},$ where $%
D=v_{F}l/3 $ is the diffusion coefficient, $v_{F}$ is the Fermi velocity and
$l=v_{F}\tau $ is the mean free path of the conduction electrons. Magnetic
impurities or a magnetic field significantly reduce the length\textbf{\ }$%
\xi _{N}$.

An impressive manifestation of the induced superconductivity in a
normal metal\textbf{\ }is the Josephson effect in S/N/S junctions.
If the thickness of the N-layer $L$ is of the order of the
correlation length $\xi _{N}$, the critical current $j_{c}$ decays
exponentially with $L$ as\textbf{\ }$j_{c}\sim \exp (-L/\xi
_{N})$\textbf{, }which means that the characteristic length of the
decay is $\xi _{N}$ and not interatomic distances. Due to this
effect the Josephson critical current can still be observed even
if the thickness of the N-layer exceeds $1$\textbf{\ }$\mu$m.

Replacing the normal metallic layer N in an S/N proximity effect
structure by a metallic ferromagnetic layer F, one comes to
basically the same effect: The pair wave function from S
penetrates into F and makes the F-layer superconducting. However,
there are important differences, rendering the S/F proximity an
interesting subject on\textbf{\ }its own.

The first important difference is that the penetration depth of
the pair wave function into the F-layer is drastically reduced as
compared to the N-layer. As will be explained in the theoretical
sections below, in the diffusive limit the penetration depth into
the ferromagnet is given by the correlation length $\xi
_{F}=\sqrt{D_{F}/2h}$ with the diffusion coefficient of the
ferromagnet $D_{F}$ and the exchange field in the ferromagnet $h$.
For strong ferromagnets like Fe, Co and Ni, the length $\xi _{F}$
has a typical value of 0.7 nm, i.e. the superconducting pairing
function decreases in F exponentially on a nearly atomic length
scale. The basic physical reason for this is that the exchange
field in the F-layer $\mathbf{h}$ tends to align the spins of a
Cooper pair and this leads to a strong pair breaking effect.

However, a faster decay of the superconducting condensate in the
ferromagnet is not the only difference in comparison with the
normal metals. Actually, there are other novel features of the S/F
proximity effect that are less obvious and they are to be the main
subject of the present review.

Since in the F-layer the spin-up and spin-down bands are split by
the exchange field $\mathbf{h}$, the electrons of a Cooper pair at
the Fermi energy have necessarily different $\mathbf{k}$-vectors
for the up and down spins and thus the Cooper pairs acquire a
finite momentum $\Delta k$. As a consequence, the condensate
function in the ferromagnet oscillates in space. As described in
the theoretical chapter below, this leads to oscillations of the
superconducting transition temperature $T_{c}$ as a function of
the F-layer thickness. For the same reason, in Josephson junctions
with an S/F/S structure, where the insulating barrier of a
conventional tunnel junction is replaced by a ferromagnetic layer,
the condensate function may change sign when crossing the F-layer,
which leads to so-called $\pi $-type coupling of
the two S-layers. This effect has been predicted long ago theoretically \cite%
{bul} but only recently confirmed experimentally
\cite{Ryazanov01,Blum}.

Following the same lines it has been demonstrated that, for a
superconducting film covered from both sides by ferromagnetic layers (F/S/F-
trilayer structure), the critical temperature $T_{c}$ depends on the mutual
orientation of the ferromagnetic layers \cite{Tagirov99,Buzdin99}. This is
the so-called superconducting spin valve effect, the latter term originating
from the possibility to switch the resistivity between the zero and a finite
values by changing the mutual magnetization direction of the two
ferromagnetic films. First experiments demonstrating this effect will be
reviewed in the subsequent experimental section.

A fascinating new aspect of S/F structures discovered recently is
a possibility of generation of a new unconventional
superconducting pairing state. The original
Bardeen-Cooper-Schrieffer (BCS) theory \cite{BCS} leads to a
conventional s-wave pairing and for several decades this type of
superconductivity has been remaining the only one observed
experimentally. On the other hand, the superconductivity in
high-$T_{c}$ cuprates discovered later shows a d-wave symmetry or
a mixture of s- and d-wave components of the order parameter
\cite{Kirtley}. Both the s-wave and d-wave types of the symmetries
of the order parameter usually imply the singlet pairing, which
means that the total spin of the Cooper pair is zero. In this
case, the order parameter $\Delta _{\alpha \beta }$ has the form
$\Delta _{\alpha
\beta }=\Delta \left( \mathbf{k}\right) \cdot (i\sigma _{2})_{\alpha \beta }$%
, where $\sigma _{2}$ is the second Pauli matrix in the spin space and $%
\Delta \left( \mathbf{k}\right) $ is a function of the momentum $\mathbf{k}$%
. As the spin part is antisymmetric with respect to transposition of the
spin indices, the antisymmetricity of the order parameter following from the
Pauli principle is fulfilled provided the function $\Delta \left( \mathbf{k}%
\right) $ is even ($\Delta \left( \mathbf{k}\right) =\Delta \left( -\mathbf{k%
}\right) $).

Another type of pairing, the spin-triplet superconductivity, has been
discovered in materials with strong electronic correlations, namely, in
heavy fermion intermetallic compounds as well as in organic materials (for a
review see \cite{Mineev}). Recently, a lot of work has been devoted to
studying superconducting properties of Sr$_{2}$RuO$_{4}$ and convincing
experimental data in favor for the triplet p-wave superconductivity have
been obtained. We refer the reader to the review articles by \cite{Maeno}
and \cite{Annett}.

In contrast to the singlet superconductivity, the spin part of the order
parameter for the triplet superconductivity is symmetric with respect to
exchanging the spin indices. Assuming that the order parameter (or the
condensate function) does not depend on frequency (which is a standard
assumption) one comes to the conclusion that the order parameter must be
antisymmetric with respect to the inversion of the momentum or,
equivalently, to transposition of the coordinates. This is this type of the
superconductivity that has been observed in, e.g., heavy fermions.

Still, there is one more, a very non-trivial possibility for triplet pairing
first predicted in Ref. \cite{BVE01b} that may be realized in S/F systems.
It turns out that the triplet pairing is also possible when the condensate
is an even function of momentum and an odd function of the Matsubara
frequency. As will be described in the theoretical sections below, a
corresponding component of the condensate function can be generated at the
S/F interface by an inhomogeneous in space magnetization. First experimental
evidence in favor of the existence of such an odd triplet superconductivity
has been reported recently \cite{Sosnin,Pena,Keizer}. A hallmark of the
triplet superconductivity in the S/F systems is its large penetration depth
in the F-layer, which follows from the fact that the exchange field in the
ferromagnet does not break the triplet pairs. Moreover, in contrast to the
\textquotedblleft conventional\textquotedblright\ triplet pairing, the odd
triplet superconductivity is not sensitive to non-magnetic disorder and
therefore is very robust.

Last but not least, in the S/F layer systems the conventional proximity
effect is not the only interesting phenomenon caused by the mutual influence
of ferromagnetic and superconducting order. As will be explained below, the
ferromagnetic state of the F-layer can in its turn be strongly modified by
the presence of the superconductor, an effect that is usually referred to as
\textit{cryptoferromagnetism} \cite{Anderson,Buzdin88,BEL}.

Remarkably, not only supercoductivity can penetrate ferromagnets but also
the S-layers can become ferromagnetic \cite{BVE04}. For the latter
phenomenon the term \textit{inverse proximity effect} was coined.

Thus, one can see that the proximity effect not only exists in the S/F
structures but there are many of them. This makes these systems extremely
interesting for both theorists and experimentalists. In the following
sections we analyze the proximity effects in the S/F systems from both
theoretical and experimental points of view. Although several reviews on the
proximity effects have been recently published \cite%
{Chien,Garifullin2002,Izyumov,Golubov04,Buzdin05,Bergeret05}, we emphasize
novel developments and open problems. At the same time, we focus mainly on
our own recent work.

\section{Proximity effect: Theory}

\subsection{S/F structures: uniform magnetization of the ferromagnet}

In this section we review shortly the main theoretical results on what
happens if one replaces the normal metal N in a N/S proximity structure by a
ferromagnetic metal F. The effective ferromagnetic exchange field acts on
spins of the conduction electrons in the ferromagnet resulting in an
additional term $\hat{H}_{ex}$ for this interaction in the total Hamiltonian
$\hat{H}_{tot}$ describing the proximity effect:
\begin{equation}
\hat{H}_{tot}=\hat{H}+\hat{H}_{ex},\ \ \ \ \ \ \ \ \hat{H}_{ex}=-\int d^{3}%
\mathbf{r}\psi _{\alpha }^{+}\left( \mathbf{r}\right) \left( \mathbf{h}%
\left( \mathbf{r}\right) \mathbf{\sigma }_{\alpha \beta }\right) \psi
_{\beta }\left( \mathbf{r}\right) d\mathbf{r},  \label{F0}
\end{equation}%
where $\psi ^{+}\left( \psi \right) $ are creation and destruction
operators, $\mathbf{h}$ is the exchange field, $\mathbf{\sigma }_{\alpha
\beta }$ are Pauli matrices, and $\alpha ,\beta $ are spin indices. The
operator $\hat{H}$ stands for a non-magnetic part of the Hamiltonian. It
includes the kinetic energy, impurities, external potentials, etc. and is
sufficient to describe all properties of the system in the absence of the
exchange field $\mathbf{h}$.

The energy of spin-up electrons differs from the energy of spin-down
electrons by the Zeeman energy $2h$. All functions, including the condensate
Green's function $f$, become matrices in the spin space with non-zero,
diagonal and off-diagonal elements. In this subsection we consider the case
of a homogeneous magnetization $M$. In this situation the matrix $\hat{f}$
is diagonal and can be represented in the form
\begin{equation}
\hat{f}=f_{3}\hat{\sigma}_{3}+f_{0}\hat{\sigma}_{0},  \label{F1}
\end{equation}%
where $f_{3}$ is the amplitude of the singlet component and $f_{0}$ is the
amplitude of the triplet component with zero projection of the magnetic
moment of the Cooper pairs on the z axis ($S=0$).{\ Note that for S/N
structures the condensate function has a singlet structure only, i.e. it is
proportional to $\mbox{$\hat{\sigma}_3$}$. The presence of the exchange
field }$\mathbf{h}$ {leads to the appearance of the triplet term
proportional to $\mbox{$\hat{\sigma}_0$}.$ }In general (non-homogeneous
magnetization) case, the matrix $\hat{f}$ contains not only the matrices $%
\hat{\sigma}_{0,3}$ but also the matrices $\hat{\sigma}_{1,2}$.

The amplitudes of the singlet and triplet components are related to the
correlation functions $\left\langle \psi _{\alpha }\psi _{\beta
}\right\rangle $ as follows \cite{Leggett,Wolfle}:
\begin{equation}
f_{0}(t)\sim \left\langle \psi _{\uparrow }(t)\psi _{\downarrow
}(0)\right\rangle +\left\langle \psi _{\downarrow }(t)\psi _{\uparrow
}(0)\right\rangle , \ \ f_{3}(t)\sim \left\langle \psi _{\uparrow }(t)\psi
_{\downarrow }(0)\right\rangle -\left\langle \psi _{\downarrow }(t)\psi
_{\uparrow }(0)\right\rangle.  \label{F2}
\end{equation}
One can see that a permutation of spins does not change the function $%
f_{3}(0),$ whereas such a permutation leads to a change of the sign of $%
f_{0}\left( 0\right) $. {This means that the amplitude of the triplet
component taken at equal times is zero in agreement with the Pauli exclusion
principle.} Later we will see that in the case of a non-homogeneous
magnetization all triplet components including $\left\langle \psi _{\uparrow
}(t)\psi _{\uparrow }(0)\right\rangle $ and $\left\langle \psi _{\downarrow
}(t)\psi _{\downarrow }(0)\right\rangle $ differ from zero.

Let us begin with a discussion of properties of the F/S systems with
homogeneous magnetization. The exchange interaction tends to align the spins
of the free electrons in one direction whereas the superconducting
correlations result in formation of Cooper pairs consisting of electrons
with opposite spins. Therefore, the superconducting transition temperature $%
T_{c}$ of an S/F bilayer system should be considerably reduced in S/F
structures provided the interface transparency is high, i.e. if the electron
can travel freely from the S to the F side and vice versa. However, it turns
out that the dependence of $T_{c}$ on the exchange field $h$ and on the
thickness of S- or F-layers is nontrivial: the critical temperature $T_{c}$
may vary with increasing $d_{F}$ in a non-monotonic way.

The critical temperature for S/F bilayer and multilayer structures was
calculated in many works \cite%
{Izyumov,Radovic91,Demler,Tagirov+C,Baladie+tc,Bagrets}. In most theoretical
papers it is assumed that the transition to the superconducting state is of
a second order, i.e. the order parameter $\Delta $ varies continuously from
zero to a finite value with decreasing temperature $T$. However, generally
this is not the case.

If the phase transition is of the second order one can linearize the
corresponding equations (the Eilenberger or Usadel equations) for the matrix
Green's function $\hat{f}$ assuming that $T$ is close to the critical
temperature $T_{c}$. This is this case that was considered in most papers on
this topic. The critical temperature of an S/F structure can be calculated
using an equation obtained from the self-consistency condition. Close to the
transition temperature $T_{c}$, the self-consistency condition can be
linearized in $\Delta ,$ and in the Matsubara representation it acquires the
form (see, e.g. Ref. \cite{Buzdin05} and references therein)
\begin{equation}
\ln \frac{T_{c}}{T_{c}^{\ast }}=(\pi T_{c}^{\ast })\sum_{\omega }(\frac{1}{%
|\omega _{n}|}-if_{\omega }),  \label{Tc}
\end{equation}%
where $T_{c}$ is the critical temperature in the absence of the proximity
effect and $T_{c}^{\ast }$ is the critical temperature when the proximity
effect is taken into account. The condensate function in Eq. (\ref{Tc}) is
the (11) element of the matrix Green's function $\hat{f}_{\omega }$. In the
diffusive case the matrix condensate function obeys the linearized Usadel
equation
\begin{equation}
D_{F}\partial ^{2}\hat{f}_{F}/\partial x^{2}-2(|\omega |\hat{\sigma}%
_{0}+ih_{\omega }\hat{\sigma}_{3})\hat{f}_{F}=0  \label{Tc1}
\end{equation}%
in the F-layer and the equation
\begin{equation}
D_{S}\partial ^{2}\hat{f}_{S}/\partial x^{2}-2(|\omega |\hat{\sigma}_{0}\hat{%
f}_{S}+i\hat{\sigma}_{3}\Delta )=0  \label{Tc2}
\end{equation}%
in the S-layer; where $D_{F,S}$ is the diffusion coefficient in the F- or
S-layer, $h_{\omega }=h\mathrm{sgn}\omega $, $h$ is the value of the
exchange field. These equations should be supplemented by boundary
conditions that near $T_{c}$ have the form (see, for example, \cite%
{Buzdin05,Bergeret05} and references therein)
\begin{equation}
\gamma _{F,S}\partial \hat{f}_{F,S}/\partial x=-(\hat{f}_{S}-\hat{f}_{F}),
\label{Tc3}
\end{equation}%
where $\gamma _{F,S}=2R_{b}\sigma _{F,S}$, $R_{b}$ is the S/F interface
resistance per unit area, $\sigma _{F,S}$ are the conductivities of the F-
and S-films in the normal state. The Usadel equation is applicable to
systems with a short mean free path $l,$ which, in other words, means that
the inverse momentum relaxation time $\tau ^{-1}$ should be larger than $%
\max \{h,2\pi T\}$ in the ferromagnet and $\tau ^{-1}$ should be larger than
$T_{c}$ in the superconductor. If these conditions are not met, one has to
solve the more complicated Eilenberger equation.

At the first glance, Eqs. (\ref{Tc1})-(\ref{Tc2}) look simple and seem to
allow a straightforward solution. However this is not so, because the order
parameter $\Delta $ depends on $x:\Delta =\Delta (x).$ In order to solve
these equations, a single-mode approximation has been introduced in several
papers \cite{Tagirov99,Izyumov,Demler,Tagirov+C}. Using this approximation
one comes to an interesting non-monotonic dependence of $T_{c}$ on $d_{F}$.
A more refined multi-mode method leads to a change of this dependence that
can be significant for some values of parameters \cite{Fominov02}.

If the interface transparency is low, Eqs. (\ref{Tc1}-\ref{Tc2}) can be
solved. The low transparency limit means that the condition $|\kappa
_{h}\gamma _{F}|<<1$ is fulfilled, where $\kappa _{h}^{2}=2(|\omega
|+ih_{\omega })/D_{F}$. In this case the condensate function in S is not
affected in the main approximation by the proximity effect and is equal to $%
\hat{f}_{S}=-i\hat{\sigma}_{3}\Delta /|\omega |$, where $\Delta $ is
approximately constant in space$.$ A solution for Eq. (\ref{Tc1}) can be
found easily
\begin{equation}
f_{\pm }(x)=\pm \frac{f_{S}}{\gamma _{F,S}\kappa _{\pm }}\exp (-\kappa _{\pm
}x),  \label{Tc5}
\end{equation}%
where $\kappa _{\pm }=\sqrt{2(|\omega |\pm ih_{\omega })/D_{F}},f_{S}=$ -$%
i\Delta /|\omega |$ and $f_{\pm }(x)$ are the $(11)$ and $(22)$ elements of
the matrix $\hat{f}_{F}$.

Usually the exchange energy $h$ is much larger than the temperature $T$ and
the expression for $\kappa _{\pm }$ shows that the condensate function $%
f_{\pm }\left( x\right) $ decays in a strong ferromagnet ($h>>T$) on a
rather short length $\xi _{F}=\sqrt{D_{F}/2h}$ and experiences oscillations
with the same period (to be more precise, the period of oscillations is $%
2\pi \xi _{F}$). This oscillatory behavior of the condensate function $f(x)$
in F leads to a non-monotonic dependence of the transition temperature $%
T_{c} $ on the thickness $d_{F}$ and to oscillations of the critical current
in S/F/S Josephson junctions (see below).

We would like to emphasize here an important point. The characteristic
length of the oscillations and decay of the condensate function is equal to $%
\xi _{F}=\sqrt{D_{F}/2h}$ only in the diffusive limit ($h\tau <<1$). In the
opposite limit ($h\tau >>1$) the situation is different: the period of the
condensate oscillations in the ferromagnet is $2\pi v_{F}/h$, whereas the
decay length is of the order of the mean free path $l$ \cite%
{Balad+Buzdin01,BVE02}.

One can easily see that the singlet component $f_{3}=(f_{+}-f_{-})/2$ is an
even function of $\omega $ and the triplet component $f_{0}=(f_{+}+f_{-})/2$
with the zero total spin ($S=0$) is an odd function of $\omega $. However,
both these components, singlet and triplet with $S=0$, coexists in the
ferromagnet over a short length of the order of $\xi _{F}$. In the next
Section we will see that in case of a non-monotonic magnetization a triplet
component with $S=\pm 1$ arises and penetrates the ferromagnet over much
larger distance of order $\xi _{N}.$

Of course, in the more complicated F/S/F structure the critical temperature
is also suppressed. However, this suppression depends also on the mutual
orientation of the magnetization in the left and right side ferromagnets.
This property has lead to the idea to switch the system between the
superconducting and normal states by varying the magnetization orientation,
Refs. \cite{Tagirov99,Buzdin99}.

Qualitatively, this effect can be understood as follows. Consider an F/S/F
structure with thin F- and S-layers. Assuming that the S/F interfaces are
highly transparent, one can average the Usadel equation over the thickness.
In this case the condensate function is continuous across the S/F
interfaces, and after averaging one can obtain an equation with an effective
order parameter $\Delta _{eff}=\Delta d_{s}/d$ and an effective exchange
field \cite{BVE01a} $h_{eff}=(h_{l}+h_{r})d_{F}/d$, where $d=2d_{F}+d_{S}$
is the total thickness of the trilayer and $h_{l}$ and $h_{r}$ are the
exchange fields from the left and from the right, respectively. Thus, the
F/S/F structure is similar to a magnetic superconductor with an effective
exchange field $h_{eff}$. It is known that the critical temperature of this
superconductor decreases with increasing $h_{eff}$ and may be even a
multivalued function of $h_{eff}$ \cite{Sarma}. If the magnetizations of the
left and right side ferromagnet are opposite to each other, we obtain $%
h_{eff}=0$ and therefore $T_{c}$ is larger than in the case of parallel
orientations of the magnetization when $h_{eff}\neq 0.$ In order to find $%
T_{c}$ in the general case, one has to solve Eqs. (\ref{Tc1}-\ref{Tc2}) with
the boundary conditions (\ref{Tc3}). These calculations have been performed
in Refs. \cite{Tagirov99,Buzdin99}.

The oscillations of the condensate function in the ferromagnet, Eq. (\ref%
{Tc5}), lead to interesting peculiarities not only in the dependence $%
T_{c}(d_{F})$ but also in the Josephson effect in the S/F/S junctions. It
turns out that under certain conditions the Josephson critical current $%
I_{c} $ changes sign and becomes negative. In this case the energy of the
Josephson coupling $E_{J}=(\hbar I_{c}/e)[1-\cos \varphi ]$ has a minimum in
the ground state when the phase difference $\varphi $ is equal not to $0$,
as in conventional Josephson junctions, but to $\pi $ (the so called $\pi -$%
junction).

This effect was predicted for the first time by Bulaevskii et al \cite{bul}%
. The authors considered a Josephson junction consisting of two
superconductors separated by a region containing magnetic impurities. Later
on, the Josephson current was calculated for a S/F/S junction \cite{Buzdin82}%
. Recently, this interesting theoretical prediction has been confirmed
experimentally \cite{Ryazanov01,Blum,Kontos02,Strunk}.

\subsection{Exotic superconductivity in S/F structures}

\label{EXO}

As mentioned in the Introduction, there is strong experimental
evidence for the realization of non-BCS type of superconducting
states in several highly correlated electron systems like the
high-$T_{c}$ cuprates, heavy fermions and Sr-ruthenates. In this
section, we will demonstrate that under certain conditions triplet
pairing is expected also in S/F systems for an arbitrary
superconductor S. In other words, the triplet component can be
artificially generated by the exchange field. Before presenting
explicit calculations, let us summarize certain general feature of
superconducting condensate using symmetry arguments.

Due to anticommutation of the fermionic creation $\psi ^{+}$ and
annihilation $\psi $ operators the condensate function $<\psi _{\alpha
}(r,t)\psi _{\beta }(r^{\prime },t^{\prime })>$ must be at equal times, $%
t=t^{\prime }$, an odd function with respect to permutations $\alpha
\leftrightarrows \beta ,$ $r\leftrightarrows r^{\prime }$. The triplet
pairing means that the spins of the Cooper pairs are parallel to each other
and the transposition of the spin indices does not change the condensate
function. Provided this function remains finite at $t=t^{\prime }$ it must
change the sign under transposition of the coordinates $r$ and $r^{\prime }$%
. So, the triplet Cooper pair has to be an odd function of the orbital
momentum or, in other words, the orbital angular momentum $L$ is an odd
number: $L=1$ (p-wave), $3$ etc.

The dependence of the condensate function on the direction in the space
makes such a superconductivity very sensitive to disorder. The p-wave
condensate (as well as d-wave pairing, etc.) is strongly suppressed by
already non-magnetic impurities. Of course, order parameter $\Delta _{\alpha
\beta }=\sum_{k}\Delta _{\alpha \beta }(\mathbf{k_{F}})\sim $ $\sum_{k}<\psi
_{\alpha }(r,t)\psi _{\beta }(r^{\prime },t)>_{k}$ is also suppressed. The
s-wave $(L=0)$ singlet condensate is an exception because it is a scalar
therefore is not destroyed by non-magnetic impurities (Anderson theorem).

At first glance, any non-singlet pairing should be suppressed by
the impurities, which makes an experimental observation very
difficult. However, one more a very non-trivial possibility for
triplet pairing exists. The previous conclusion about the
antisymmetricity of the orbital part of the condensate function
remains finite at equal times, which excluded functions
antisymmetric in $t-t^{\prime }$.

At the same time, nothing forbids the function $<\psi _{\alpha }(r,t)\psi
_{\beta }(r^{\prime },t)>$ to change sign under the transposition $%
t\leftrightarrows t^{\prime }$. In the frequency representation, this
property is realized if the correlator $<\psi _{\alpha }(r,\tau )\psi
_{\beta }(r^{\prime },\tau ^{\prime })>_{k,\omega }$ is an odd function of
the Matsubara frequency $\omega .$ However, if the condensate function is
odd in frequency, it may be even in the momentum and we come to the triplet
pairing again. In this case, the correlation function $<\psi _{\alpha
}(r,\tau )\psi _{\beta }(r^{\prime },\tau ^{\prime })>_{k,\omega }$equals
zero at coinciding times (the sum over all frequencies is zero) and
therefore the Pauli principle for the equal-time correlators is not violated.

This type of pairing was suggested by Berezinskii \cite{berezin} as a
possible mechanism of superfluidity of $^{3}$He. He assumed that the order
parameter $\Delta (\omega )\propto \sum_{\omega ,k}<\psi _{\alpha }(r,\tau
)\psi _{\beta }(r^{\prime },\tau ^{\prime })>_{k,\omega }$ is an odd
function of $\omega :$ $\ \Delta (\omega )=-\Delta (-\omega )$. However
experiments on superfluid $^{3}$He have shown that the Berezinskii's state
was not realized in this system. Now it is well known that the condensate in
$^{3}$He is antisymmetric in the momentum space and symmetric (triplet) in
the spin space. Thus, the Berezinskii hypothetical pairing mechanism
remained unrealized for few decades.

Recent theoretical studies have shown that a superconducting state
similar to the one suggested by Berezinskii might be induced in
S/F systems due to the proximity effect \cite{BVE01b,BVE03}. {In
the next sections we will analyze this new type of
superconductivity with triplet pairing that is odd in frequency
and, in the diffusive limit, even in momentum}. This can be s-wave
pairing and therefore this type of the superconductivity is not
sensitive to impurities.

We note, however, that there is a qualitative difference between this new
superconducting state in S/F structures and the one proposed by Berezinskii.
In S/F structures both singlet and triplet types of the condensate coexist
and the order parameter $\Delta $ existing only in the S region (we assume
that the superconducting coupling in the F region is zero) is determined
solely by the singlet part of the condensate.

Note that, while theories of unconventional superconductivity often imply
strongly correlated systems, the triplet state induced in S/F structures can
be derived within the framework of the BCS theory valid in the weak-coupling
limit. This fact not only drastically simplifies theoretical considerations
but also helps in designing experiments, since well known elemental
superconductors prepared under controlled growth procedures may be used in
order to detect the triplet superconductivity.

To finish this subsection let us summarize the properties of this new type
of superconductivity that we call \textit{odd} \textit{triplet
superconductivity}:

\begin{itemize}
\item {It contains a triplet component. In particular the components with
projection $S=\pm 1$ are insensitive to the presence of an exchange field
and therefore long-range proximity effects arise in S/F structures.}

\item {In the dirty limit it has an s-wave symmetry. The condensate function
is even in $\mathbf{p}$ and therefore, contrary to the unconventional
superconductors with triplet pairing, is not destroyed by the presence of
non-magnetic impurities. }

\item {The triplet condensate function is \textit{odd} in frequency}.
\end{itemize}

Before we turn to a more detailed theoretical analysis of the triplet
superconductivity, we remark that in the F-regions of the S/F structures no
attractive electron-electron interaction exists, and therefore $\Delta =0$
in F. This means that only the superconducting condensate function $f$ in
the ferromagnet exists and, as it will become clear later, it arises only
due to the proximity effect.

\subsection{Triplet Odd Superconductivity induced by an inhomogeneous
magnetization in S/F structures}

\label{EXO-INH}

As discussed in 2.1, the presence of an exchange field results in formation
of the triplet component of the condensate function. In a homogeneous
exchange field, only the component with the projection $S=0$ is induced.
Then the natural question arises: Can the other components with $S=\pm 1$
also be induced? If they could, this would lead to a long range penetration
of the superconducting correlations into the ferromagnet because these
components correspond to the correlations of the type $<\psi _{\uparrow
}\psi _{\uparrow }>$ with parallel spins and they are not as sensitive to
the exchange field as the other ones.

In what follows, we analyze a few examples of the S/F structures in which
all the projections of the triplet component are induced. The common feature
of these structures is that the magnetization should be non-homogeneous.

\subsubsection{F/S/F trilayer structure}

\label{EXO-INH-FSF} We start with considering the F/S/F system shown
schematically in Fig. 1. The structure consists of one S-layer and two
F-layers with the magnetization inclined at the angle $\pm \alpha $ with
respect to the $z$-axis (in the $yz$ plane).
\begin{figure}[th]
\centering{\includegraphics[width=6cm]{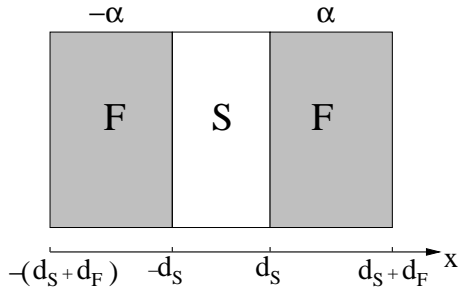}}
\caption{Trilayer geometry. The magnetization of the left (right) side
F-layer makes an angle $\protect\alpha $ ($-\protect\alpha $) with the $z$%
-axis.}
\end{figure}
As we have seen in the previous section, each of the layers generates the
triplet component with the zero total projection of the spin, $S=0,$ in the
direction of the exchange field. If the magnetic moments of the layers are
collinear (parallel or antiparallel), the total projection remains zero.
However, if the moments of the ferromagnetic layers are not collinear, the
superposition of the triplet components coming from the different layers
should have all the possible projections of the total spin.

From this qualitative argument we can really expect the non-trivial effect
of the generation of the triplet components with all the projections of the
total spin provided\ the thickness of the S-layer is not too large. The
point is that the triplet component decays in S on a length of the order of
the coherence length $\xi _{S}\approx \sqrt{D_{S}/\pi T_{c}}$, Eq. (\ref{Tc2}%
). We assume that the thickness of the S-layer does not exceed much this
length.

In order to find all types of the condensate (singlet and triplet), one has
to solve the linearized Usadel equation in the F-region (we assume a weak
proximity effect) \cite{BVE03} for the condensate function $\check{f}$ that
is a $4\times 4$ matrix in the particle-hole and spin spaces
\begin{equation}
(\partial ^{2}\check{f}/\partial x^{2})-\kappa _{\omega }^{2}\check{f}%
+i\kappa _{h}^{2}\{\hat{\tau _{0}\otimes }[\hat{\sigma}_{3},\check{f}%
]_{+}\cos \alpha \pm \hat{\tau _{3}\otimes }[\hat{\sigma}_{2},\check{f}]\sin
\alpha \}=0\;,  \label{FSF1}
\end{equation}%
where $[\hat{\sigma}_{3},\check{f}]_{+}=\hat{\sigma}_{3}\otimes \check{f}+%
\check{f}\otimes \hat{\sigma}_{3}$.

The wave vectors $\kappa _{\omega }$ and $\kappa _{h}$ entering Eq. (\ref%
{FSF1}) have the form
\begin{equation}
\kappa _{\omega }^{2}=2|\omega |/D_{F},\quad \kappa _{h}^{2}=2h\mathrm{sgn}%
(\omega )/D_{F}  \label{a1}
\end{equation}

The magnetization vector $\mathbf{M}$ lies in the ($y,z$)-plane and has the
components: $\mathbf{M}=M\{0,\pm \sin \alpha ,\cos \alpha \}$. The sign
\textquotedblleft +\textquotedblright\ (\textquotedblleft
-\textquotedblright ) corresponds to the right (left) side F-film. We
consider here the simplest case of a highly transparent S/F interface and
temperatures close to the superconducting transition temperature $T_{c}$. In
this case the function $\check{f}$, being small, obeys a linear equation in
S {similar to Eq. (\ref{Tc2}).}

The boundary conditions at the S/F interfaces are obtained by a
generalization of Eqs. ({\ref{Tc3}}) (see Ref. \cite{BVE03}).

A solution for Eqs. (\ref{FSF1}) can easily be found. The matrix $\check{f}$
can be represented as
\begin{equation}
\check{f}=i\hat{\tau}_{2}\otimes \hat{f}_{2}+i\hat{\tau}_{1}\otimes \hat{f}%
_{1},  \label{FSF3}
\end{equation}
where $\hat{f}_{1}=b_{1}(x)\hat{\sigma}_{1},$ \ $\hat{f}_{2}=b_{3}(x)\hat{%
\sigma}_{3}+b_{0}(x)\hat{\sigma}_{0}$.

For the left side F-layer the functions $b_{k}(x)$ are to be replaced by $%
\bar{b}_{k}(x).$ For simplicity we assume that the thickness of the F-films $%
d_{F}$ exceeds $\xi _{F}$ (the case of an arbitrary $d_{F}$ was analyzed in
\cite{BVE03}). Using the representation, Eq. (\ref{FSF3}), we find the
functions $b_{i}(x)$ and $\bar{b}_{i}(x)$. They are decaying exponential
functions and can be written as
\begin{equation}
b_{k}(x)=b_{k}\exp (-\kappa (x-d_{S})),\quad \bar{b}_{k}(x)=\bar{b}_{k}\exp
(\kappa (x+d_{S})).  \label{FSF4}
\end{equation}%
Substituting Eq. (\ref{FSF4}) into Eq. (\ref{FSF1}), we obtain a set of
linear equations for the coefficients $b_{k}$ that should be complemented by
expressions for the eigenvalues $\kappa .$

In the limit of large exchange energy $h$ ($\{T,\Delta \}<<h,$ but $h<<\tau
^{-1}$), the eigenvalues $\kappa $ are equal to
\begin{equation}
\kappa =\kappa _{\omega },\ \ \ \kappa _{\pm }\approx (1\pm i)\kappa _{h}
\label{FSF5}
\end{equation}

We see from Eq. (\ref{FSF5}) that the solutions $\kappa _{\omega }$ and $%
\kappa _{\pm }$ are completely different. {The roots $\kappa _{\pm }$
proportional to }$\kappa _{h}$ are very large and therefore the
corresponding solutions $b_{k}\left( x\right) $ decay very fast (similar to
the singlet component). This is the solution that exists for a homogeneous
magnetization (collinear magnetization vectors).

In contrast to the roots $\kappa _{\pm }$, the value for $\kappa
_{\omega }$ given by Eq. (\ref{FSF5}) does not depend on the
exchange energy $h\ $and is much smaller. It is this eigenvalue
that leads a to slow decay of the
superconducting correlations. The solutions corresponding to the root $%
\kappa _{\omega }$ describe the long-range penetration of the triplet
component into the ferromagnetic region. The function $b_{1}(x)$ is the
amplitude of the triplet component penetrating the F-region over a long
distance of the order of $\kappa _{\omega }^{-1}\sim \xi _{N}.$ Its value as
well as the values of the other functions $b_{k}(x)$ is to be found from the
boundary conditions at the S/F interfaces.

Matching the solutions in S and F at the S/F interfaces, we obtain the
coefficients $b_{k}$ and $\bar{b}_{k}$. Note that $b_{3\pm }=$ $\bar{b}%
_{3\pm }$ and $b_{\omega }=$ $-\bar{b}_{\omega }.$ Although the solution can
be found for arbitrary parameters entering the equations, we present here
the expressions for $b_{3\pm }$ and $b_{\omega }$ in some limiting cases
only. For example, if the parameter $\gamma \kappa _{h}/\kappa _{S}$ is
small, the amplitudes of the long-range triplet component $b_{\omega }$ and
singlet components $b_{3\pm }$ can be written in a rather simple form
\begin{equation}
b_{\omega }\approx -\frac{2\Delta }{E_{\omega }}(\frac{\gamma \kappa _{h}}{%
\kappa _{S}})\frac{\sin \alpha \cos ^{2}\alpha }{\sinh (2\Theta _{S})},\ \
b_{3+}\approx b_{3-}\approx \frac{\Delta }{2iE_{\omega }}  \label{FSF6}
\end{equation}%
where\textbf{\ }$\Theta =\kappa _{S}d_{S}$ and $E_{\omega }=\sqrt{\omega
^{2}+\Delta ^{2}},$ $\gamma =\sigma _{F}/\sigma _{S}$ and $\sigma _{F}\left(
\sigma _{S}\right) $ is the conductivity in the ferromagnet (superconductor).

At the S/F interface the amplitude of the triplet component $b_{\omega }$ is
small compared to the magnitude of the singlet one $b_{3+}$. However, the
triplet component decays over a long distance $\xi _{N}$, while the singlet
one vanishes at distances exceeding the short length $\xi _{F}$. The
amplitudes $b_{\omega }$ and $b_{3\pm }$ become comparable if the parameter $%
\gamma \kappa _{h}/\kappa _{S}$ is of the order of unity.

It follows also from Eq. (\ref{FSF6}) that the amplitude of the triplet
component $b_{\omega }$ is zero in the case of collinear vectors of
magnetization, i.e. at $\alpha =0$ or $\alpha =\pi /2$. It reaches the
maximum at the angle $\alpha _{m}$ for which $\sin \alpha _{m}=1/\sqrt{3}$.
Therefore the maximum angle-dependent factor in Eq. (\ref{FSF6}) is $\sin
\alpha _{m}\cos ^{2}\alpha _{m}=2/3\sqrt{3}\approx 0.385.$

One can see from Eq. (\ref{FSF6}) that $b_{\omega }$ becomes exponentially
small if the thickness $d_{S}$ of the S-films significantly exceeds the
coherence length $\xi _{S}\approx \sqrt{D_{S}/\pi T_{c}}$. This means that
in order to have a considerable penetration of the superconducting
condensate into the ferromagnet, one should not make the superconducting
layer too thick. On the other hand, if the thickness $d_{S}$ {is too small},
the critical temperature $T_{c}$ is suppressed. In order to avoid this
suppression, one has to use, for instance, a F/S/F structure with a small
thickness of the F-films. In Fig. 2 we plot the spatial dependence of the
singlet and triplet components in F/S/F structure.
\begin{figure}[th]
\centering{\includegraphics[width=6cm]{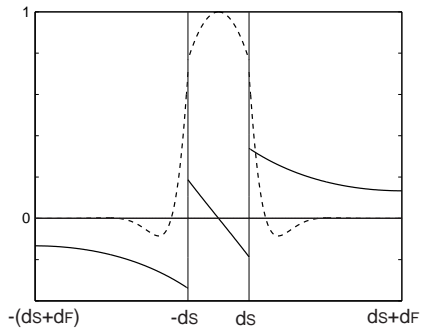}}
\caption{The spatial dependence of Im($b_{3}(x)$) (dashed line) and the
long-range part of Re($b(x)$) (solid line). We have chosen $\protect\sigma %
_{F}/\protect\sigma _{S}=0.2,h/T_{c}=50,\protect\sigma _{F}R_{b}/\protect\xi %
_{F}=0.05,d_{F}\protect\sqrt{T_{c}/D_{S}}=2,d_{S}\protect\sqrt{T_{c}/D_{S}}%
=0.4$ and $\protect\alpha =\protect\pi /4$. The discontinuity of the triplet
component at the S/F interface is because the short-range part is not shown
in this figure. Taken from \protect\cite{BVE03}. }
\end{figure}

\subsubsection{Domain wall at the S/F interface and helical ferromagnets}

\label{EXO-INH-DW} Now we consider another example of an S/F
structure in which the long-range triplet component (LRTC) also
arises. This structure is shown schematically in Fig. 3.
\begin{figure}[th]
\centering{\includegraphics[width=6cm]{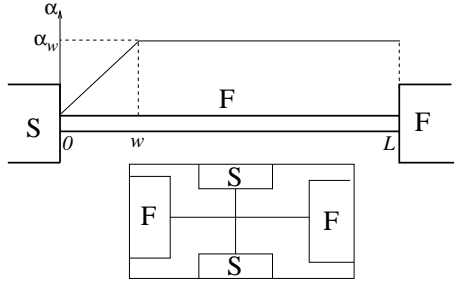}}
\caption{S/F structure with a domain wall in the region $0<x<w$. In this
region $\protect\alpha =Qx$, where $Q$ is the wave vector which describes
the spiral structure of the domain wall. For $x>w$ it is assumed that the
magnetization is homogeneous, i.e., $\protect\alpha =Qw$.}
\end{figure}
It consists of an S/F bilayer with a non-homogeneous magnetization
in the
F-layer. We assume for simplicity that the magnetization vector $\mathbf{M}%
=M(0,\sin \alpha ,\cos \alpha )$ rotates in the F-film starting from the S/F
interface ($x=0$), and the rotation angle has a simple, piece-wise $x$%
-dependence: $\alpha (x)=Qx,$ in the domain wall for\ $0<x<w$ and $\alpha
(x)=Qw$ for $w<x$. {This means} that the $\mathbf{M}$ vector is directed
parallel to the $z$-axis at the S/F interface and rotates by the angle $%
\alpha (w)$ over the length $w$ ($w$ may be the width of a domain wall). At $%
x>w$ the orientation of the vector $\mathbf{M}$ is fixed.{\ This structure
was considered first in }\cite{BVE01b} and later in \cite{Kadig}. The Usadel
equation for this case has been solved in Ref. \cite{BVE01b}. The solution
is found in the region of the domain wall $0<x<w$ and in the region of a
constant magnetization: $w<x<\infty $.

For this configuration of the magnetic moment the LRTC arises in the domain
wall and spreads into the ferromagnet over a long distance. The
characteristic decay length of the LRTC inside the domain wall is
\begin{equation}
\xi _{Q}=(Q^{2}+\kappa _{\omega }^{2})^{-1/2}  \label{Dom2}
\end{equation}%
whereas its value equals $\kappa _{\omega }^{-1}$ outside the
domain wall. The singlet component penetrates the ferromagnet over
a short length of the order of $\xi _{F}.$ Although the amplitude
of the LRTC at the S/F interface may be comparable with the
amplitude of the singlet component, the decay length of the LRTC
is much larger (see Fig. 4).
\begin{figure}[th]
\centering{\includegraphics[width=6cm]{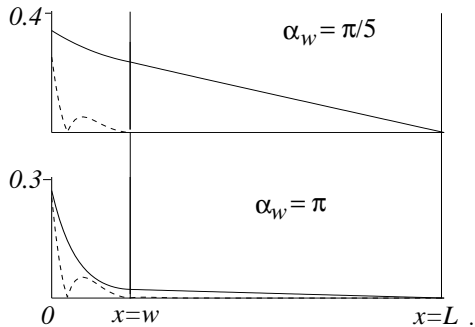}} .
\caption{Spatial dependence of amplitudes of the singlet (dashed line) and
triplet (solid line) components of the condensate function in the F wire for
different values of $\protect\alpha _{w}$. Here $w=L/5,\protect\epsilon %
=E_{T}$, and $h/E_{T}=400$. $E_{T}=D_{F}/L^{2}$ is the Thouless energy,$%
\protect\epsilon =i\protect\omega $ is the energy (From \protect\cite{BVE01b}%
)}
\end{figure}
One more system where the LRTC arises is a helical ferromagnet \cite{VAE}
(see Fig. 15). Such a structure is realized, for example, in several heavy
rare earth metals. In this ferromagnet the magnetization vector rotates
around the $z$-axis and has a non-zero projection $H_{z}$ on this axis.

It was shown that in this case the LRTC penetrates the ferromagnet over a
length of the order of $\xi _{Q},$ Eq. (\ref{Dom2}). What is interesting,
the monotonic decay of the LRTC in this case occurs only if the the cone
angle $\theta $ is less than $\sin ^{-1}(1/3)\approx 19^{o}$. At larger $%
\theta $ the decay of the LRTC is accompanied by oscillations. In the
quasi-ballistic case ($h\tau >1$), the characteristic length of the LRTC
penetration into the ferromagnet changes.

In the case of Neel-type domain walls the LRTC vanishes provided
the magnetization vector $M$ rotates continuously \cite{Eschrig2}.
However, in an S/F structure with several Neel domain walls (the
vector $M$ rotates only inside the domain walls) the LRTC arises
at the domain walls and decays in the domains over a large
distance \cite{Fominov2}.

\subsection{Other proximity effects in S/F structures}

\label{RED}Up to now we implicitly assumed that the proximity effect in S/F
structures changes the superconducting properties but leaves the
magnetization of the F-layer unchanged. However, this is not always true and
experiments performed by \cite{Muhge98} and \cite{Garifullin2} indicate that
the ferromagnetic magnetization of S/F bilayers may decrease when lowering
the temperature below the superconducting transition temperature $T_{c}$. At
that time it was not quite clear what physical mechanism causes this
decrease of the magnetization. Here we review two different and independent
mechanisms that may explain the effect.

\subsection{Cryproferromagnetism}

In a classic paper Anderson and Suhl \cite{Anderson} proposed an
idea that at some circumstances superconductivity might coexist
with a non-homogeneous magnetic ordering. They called this
magnetic non-homogeneously ordered state
\textit{cryptoferromagnetic}. The basic reasoning leading to this
suggestion was that superconductivity could survive in a
ferromagnetic background, if the magnetization direction varied on
a scale smaller than the superconducting coherence length. The
cryptoferromagnetic state in S/F structures was considered first
in Ref. \cite{Buzdin88} in the case of a weak ferromagnet.

In a more recent theoretical paper on cryptoferromagnetism in S/F bilayers
\cite{BEL} a more realistic case of a strong ferromagnet was considered. It
was shown that even if the exchange field is large the cryptoferromagnetic
state is still possible provided the ferromagnetic film is sufficiently
thin. A phase diagram containing the cryptoferromagnetic state has been
drawn depending on the stiffness of the ferromagnet $\mathcal{J}$ , the
thickness of the F-film $d_{F}$ and the exchange field $h$ of the system.
This phase diagram ($a,\lambda $) for the S/F system is represented in Fig.
5, where $a=2h^{2}d_{F}^{2}/(D_{F}T_{c}\eta ^{2}),\eta =v_{F}/v_{S},$ $%
\lambda =(\mathcal{J}d_{F}/N_{F}\sqrt{2T_{c}D_{F}^{3}})(7\varsigma (3)/2\pi
^{2}),$ and $N_{F}$ is the density-of-states (DOS) in the ferromagnet.
\begin{figure}[th]
\centering{\includegraphics[width=9cm]{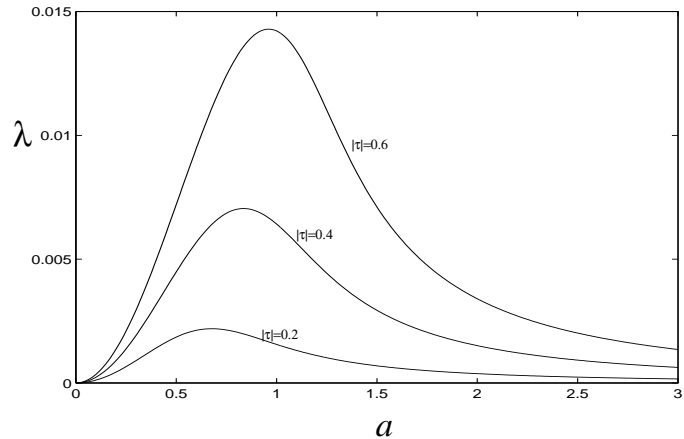}}
\caption{Phase diagrams ($\protect\lambda ,\protect\alpha $) for different
values of $|\protect\tau |=(T_{c}-T)/T_{c}$. The area above (below) the
curves corresponds to the F (CF) state.}
\end{figure}
Estimates of the parameters ($\mathcal{J}$, $h$ and $d_{F}$) for the samples
used in the experiments \cite{Garifullin2} in which a reduction of the
effective magnetization was observed show that the results of Ref. \cite{BEL}
agree with the experimental data.

The calculations show that the proximity effect may lead to a magnetic
spiral structure in the F-film even if the exchange energy $h$\ is much
larger than the characteristic energy of the superconducting transition $%
T_{c}$. This cryptoferromagnetic ordering is related to existence of low
lying states in the ferromagnet. The spiral structure increases the magnetic
energy only by a small amount, whereas the energy of interaction between the
exchange field and the superconductivity can essentially be reduced.

At the same time, there exists another mechanism that can reduce the total
magnetization in S/F structure and it is also due to the proximity effect.
This is the so-called inverse proximity effect describing the situation when
the orientation of the magnetization remains unchanged, while its magnitude
changes both in the S- and F-layers.

\subsection{Inverse proximity effect}

\label{RED-INV} The inverse proximity effect is due to a
contribution of free electrons both in the ferromagnet ($\delta
M_{F}$) and in the superconductor\ ($M_{S}$) to the total
magnetization. On one hand, the DOS in the F-film is reduced due
to the proximity effect, thus decreasing the magnetization in F by
$\delta M_{F}$. On the other hand, the Cooper pairs in S are
polarized in the direction opposite to $M_{F}$, giving rise to a
magnetization ($M_{S}$) with a direction opposite to $M_{F}$. So,
the S-layer becomes ferromagnetic and this is the reason for
calling this effect the inverse proximity effect.
\begin{figure}[th]
\centering{\includegraphics[width=7cm]{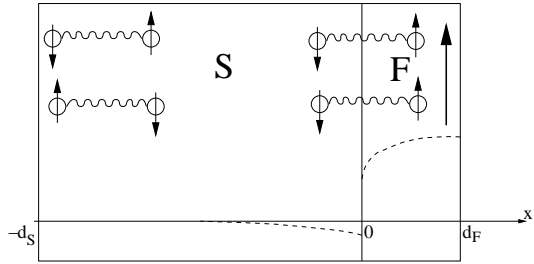}}
\caption{S/F structure and schematic representation of the inverse proximity
effect. The dashed curves show the local magnetization.}
\end{figure}
For a more detailed qualitative explanation of this mechanism we consider
the S/F structure with a thin F-layer in Fig. 6. We assume that the exchange
field of F is homogeneous and directed along the $z$-axis.

If temperature exceeds the critical temperature $T_{c}$, the total
magnetization of the system $M_{tot}$\ equals $M_{0}d_{F}$, where $d_{F}$\
is the thickness of the F-layer. When the temperature is lowered below $%
T_{c} $, the S-layer becomes superconducting and Cooper pairs with a size of
the order of $\xi _{S}$ appear in the superconductor. Due to the proximity
effect the Cooper pairs cross the interface and penetrate into the
ferromagnet. In the case of a homogeneous magnetization in F the Cooper
pairs are composed of electrons with opposite spins, such that the total
magnetic moment of the pair equals zero. The exchange field is assumed to be
not too strong, otherwise the Cooper pairs would be destroyed.

It is clear from this simple picture that pairs located entirely in the
superconductor cannot contribute to the magnetic moment of the
superconductor. However, some pairs are distributed in space in a more
complicated manner: one of the electrons of the Cooper pair stays in the
superconductor, whereas the other one enters the ferromagnet. These are
these pairs that create the magnetic moment in the superconductor.

Energetically it is favorable for the electron of the Cooper pair with the
spin parallel to the magnetization of the ferromagnet to have a higher
probability density in F. This means that the electron with the opposite
spin has a higher probability density in S. This is the reason why these
pairs form a magnetic moment in the S-layer. As a result, the ferromagnetic
order is created in the S-layer with a direction of the magnetic moment
opposite to the direction of $\mathbf{M}$ in F-layer. The induced magnetic
moment penetrates the superconductor over the size of the Cooper pairs,
which may be much larger than $d_{F}$.

Using similar arguments we can predict a related effect: the magnetic moment
in the ferromagnet should be reduced in the presence of superconductivity
because some of the electrons located entirely in the ferromagnet condense
into Cooper pairs and do not contribute to the magnetization.

From this qualitative, simplified picture one can expect that the total
magnetization of an S/F system will be reduced for temperatures below $T_{c}$%
. A quantitative analysis based on the Usadel equation (diffusive case) \cite%
{BVE04} or on the Eilenberger equation (quasiballistic case) \cite%
{Kharitonov} supports the qualitative picture. It turned out that at low
temperatures the magnetic moment $M_{F}$ in F is screened completely by the
spin-polarized Cooper pairs in S if $M_{F}$ is due to free electrons (ideal
itinerant ferromagnet) i.e. $M_{S}=-M_{F}$. This conclusion is valid in the
limit $h<$ $D_{F}/d_{F}^{2}$.

With increasing the exchange energy $h$ the induced magnetic moment
decreases monotonically in the diffusive limit \cite{BVE04} or
non-monotonically in the clean limit \cite{Kharitonov}.

It should be stressed that both the mechanism discussed here and that of the
last section lead to a decrease of the total magnetization. The spin
polarization of Cooper pairs in the superconductor in F/S/F structure with a
non-collinear magnetization in F was studied in Ref. \cite%
{Lofwander}.

\section{S/F proximity effect: Experiments}

\subsection{Superconducting transition temperature in F/S systems}

Following the theoretical predictions (see Sec.2.2) a
ferromagnetic film deposited on a superconducting film should
drastically suppress the superconducting $T_{c}$. In experimental
systems, however, this is often not the case, the $T_{c}$-
suppression appears rather moderate. This is due to two different
reasons. First, in real thin film systems there is often an
intermediate alloy layer caused by interdiffusion that is weakly
magnetic or even non-magnetic. This is the case, e.g., in Fe/Nb
\cite{Muhge96,Muhge97} and, probably, in Gd/Nb \cite{Jiang}. This
interlayer prevents the direct contact between the F- and S-layer
and weakens the suppression of $T_{c}$. Second, the quantum
mechanical transparency of a real S/F interface is often quite
small, i.e. the coefficient $\gamma _{F,S}$ in Eq. (8) is small
and the $T_{c}$-suppression is much weaker than that with an
ideally transparent interface.

An interesting feature of the S/F proximity effect that has recently been
under intensive discussion in literature, is oscillation of the
superconducting transition temperature as a function of the F-layer
thickness $d_{F}$.

There are quite different physical mechanisms that may cause $T_{c}(d_{F})$
oscillations or a non monotonic $T_{c}(d_{F})$ behavior. An indirect
mechanism, not directly related to the proximity effect, has been observed
in Fe/Nb bilayers \cite{Muhge96,Muhge97} (Fig. 7c). Here an alloying at the
interface leads to a non-ferromagnetic NbFe interlayer of about 0.7 nm
thickness and therefore the minimum in $T_{c}(d_{F})$ just correlates with
the onset of ferromagnetism.
\begin{figure}[th]
\centering{\includegraphics[width=7cm]{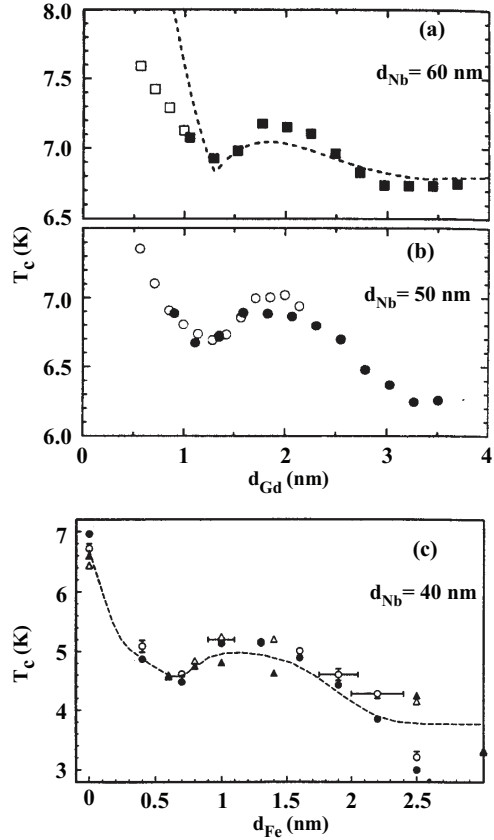}}
\caption{Dependence of superconducting transition temperature on the
ferromagnetic layer thickness in (a and b) two series of Nb/Gd multilayers (
Jiang et al \protect\cite{Jiang}), and (c) Fe/Nb/Fe trilayers (M{\"{u}}hge
et al \protect\cite{Muhge96,Muhge97}).}
\end{figure}
The explanation of this phenomenon is that strong longitudinal spin
fluctuations exist in the NbFe interlayer with a concentration close to the
onset of ferromagnetic long range order. They are responsible for the strong
initial $T_{c}$-suppression when increasing the Fe-thickness from 0 to 0.7
 nm in Fig. 7c. When the first ferromagnetic Fe layer appears above $d_{Fe}$\
=0.7 nm, the spin fluctuations in the NbFe interlayer are suppressed by the
exchange field of the Fe-layer and result in an increase of $T_{c}$. This is
a rather indirect influence of the ferromagnetic Fe-layer on the
superconductivity of the Nb layer.

Now, coming to oscillations in $T_{c}(d_{F})$ induced by the S/F proximity
effect, we recall that, as discussed in Sec. 2.2, oscillations of the
condensate function $\hat{f}_{s}$ in space may directly lead to a
non-monotonic $T_{c}$\ -dependence on $h$ or $d_{F}$. Actually the reason
for this non monotonic behavior may be different for S/F bilayers and S/F/S
trilayers, since the boundary conditions in these two cases are different.
For bilayers, only one side of the F-layer is in contact with the
superconductor, whereas in the S/F/S trilayers the F-layer is in contact
with the superconducting layers on the both sides.

For the case of trilayers (but not for bilayers) oscillations of $T_{c}$\
may be due to appearance (or disappearance) of Josephson $\pi $-coupling. As
mentioned in Sec. 2.3, due to the oscillation of the superconducting pairing
function in the F-layer, the phase difference in the superconducting pairing
function on both the sides of the F-layer may have opposite phases at
certain F-layer thicknesses, i.e., the phase difference between the
neighboring S-layers may be equal to $\pi $. Radovi\'{c} et al \cite%
{Radovic91} concluded from their calculations that $T_{c}$ for $\pi $%
-coupling most probably is higher than for the vanishing phase difference.

Jiang et al \cite{Jiang} claimed that the observed oscillations of $%
T_{c}(d_{Gd})$ (Fig. 7a and 7b) are due to this type of the Josephson $\pi $%
-coupling. Several other works on S/F multilayers have reported a single
peak in $T_{c}(d_{F})$, and have attributed this feature to "$\pi $%
-switching" (see, e.g., \cite{Jiang96,Chien97,Ogrin00}).

Whereas the role of $\pi $-coupling for the non monotonous $T_{c}(d_{F})$ in
S/F multilayers has not been finally clarified and an alternative
explanation exists (see below), a clear experimental evidence for $\pi $%
-coupling across an F-layer comes from the study of Josephson junctions
using F-layers as barriers \cite{Ryazanov01,Kontos02}. Tunnelling
spectroscopy revealed damped oscillations of the superconducting order
parameter induced in the F-film by the proximity effect \cite{Kontos02}.
Ryazanov et al \cite{Ryazanov01} performed measurements of the critical
current in Josephson junctions consisting of superconducting Nb and weakly
ferromagnetic interlayers and found that the character of the junction
changed from $0$-phase at high temperatures to $\pi $-phase at low
temperatures. This result has later been confirmed by Blum et al \cite{Blum}%
. A different phase sensitive experiment \cite{Guichard} also gave evidence
for the oscillatory behavior of the critical supercurrent of S/F layered
system when varying the F-layer thickness.

Without invoking the $\pi $-coupling, oscillations of $T_{c}(d_{F})$ can
simply originate from the oscillations of the condensate amplitude in space
within the F-layer. As shown theoretically by \cite%
{Radovic91,Tagirov+C,Proshin}, due to these oscillations and taking the
boundary conditions for the pairing wave function at the S/F interfaces into
account, the $T_{c}(d_{F})$-curve may have an oscillatory character with the
oscillation period of the order $\xi _{h}=v_{F}/h$ (see Sec. 2.2).

The physical origin of the oscillatory character of $T_{c}(d_{F})$ can
qualitatively be traced back to the propagating character of the
superconducting pairing wave function in the ferromagnet. If the thickness
of the F-layer is smaller than the penetration depth of the pairing wave
function, this function, when transmitted through the S/F interface into the
F-layer, will interfere with the wave reflected from the opposite surface of
the ferromagnet. As a result, the flux of the pairing wave function crossing
the S/F interface varies with the thickness of the F-layer $d_{F}$.

Then, the coupling between the electrons of the ferromagnet and the
superconductor will be modulated and $T_{c}$ will oscillate with $d_{F}$. If
the interference at the S/F interface is essentially constructive (this
corresponds to a minimal jump of the pairing function amplitude at the S/F
interface), the coupling is weak, and one expects $T_{c}$ to be maximal.
When the interference is destructive, the coupling is maximized and $%
T_{c}(d_{F})$ is minimal.

It should be noted that this model explaining $T_{c}(d_{F})$ oscillations
applies to the case of the S/F bilayers as well as to F/S/F trilayers or S/F
multilayers, whereas the $\pi $-coupling concept does not apply for bilayers.

Aarts et al \cite{Aarts} studied V/V$_{1-x}$Fe$_{x}$ multilayers without
interdiffusion at the interface. They showed that $T_{c}$ strongly depends
on the interface transparency and presented experimental evidence for an
intrinsically reduced interface transparency. From the dependence of $T_{c}$
on the magnetic layer thickness they calculated the penetration depth of
Cooper pairs into the F-layer and found it to be inversely proportional to
the effective magnetic moment per Fe atom. For the interpretation of the
observed peculiarities they introduced a finite transparency of the S/F
interface and argued, based on their experimental data, that with an
increasing the exchange splitting of the conduction band in the F-layer the
transparency of the S/F interface for Cooper pairs decreases.

Lazar et al \cite{Lazar} studied experimentally the role of the interface
transparency in the Fe/Pb/Fe system and, for comparison, in the Fe/V/Fe
system, too \cite{Garifullin02}. In contrast to the case of Fe/Nb/Fe
discussed above, in Fe/Pb/Fe and Fe/V/Fe the intermixing at the interfaces
is much weaker. Fig. 8 shows the dependence of $T_{c}$ on the thicknesses of
the Fe layers for Fe/Pb/Fe trilayers.
\begin{figure}[th]
\centering{\includegraphics[width=4cm]{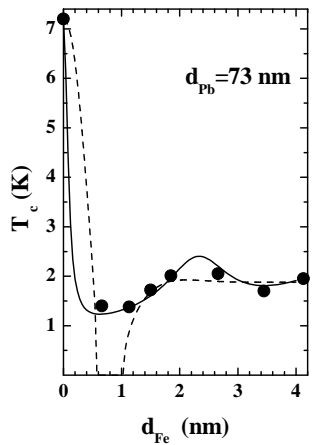}}
\caption{$T_{c}$ dependence on the Fe thickness at fixed value $d_{Pb}=73$
 nm for Fe/Pb/Fe trilayers. The dashed line is obtained by the Radovi\'{c} et
al \protect\cite{Radovic91} theory which supposes an ideally transparent
interface. The solid line takes a finite transparency of the interface into
account \protect\cite{Tagirov+C}.}
\end{figure}
A theoretical analysis of the curves using model calculations revealed that
the experimental results can only be described assuming Pb/Fe interfaces
that are not perfectly transparent. The critical temperature $T_{c}$ for the
case of S/F interfaces with a non-perfect transparency has been calculated
by Golubov \cite{Aarts} and Tagirov \cite{Tagirov+C}.

A fit to the experimental points using the model calculations \cite%
{Tagirov+C} is plotted on Fig. 8 as a solid line. The quality of the fit is
satisfactory and reproduces the details of the $T_{c}(d_{Fe})$-curve. The
most important parameter obtained from this fit is the value of $T_{m}$,
characterizing the transparency of the interface. The fit gives $T_{m}$=0.4.
This value corresponds to a quantum mechanical transmission coefficient $%
\overline{T}=T_{m}/(1+T_{m})=0.3$ \cite{Lazar} that is considerably reduced
as compared to the ideally transparent interface with $\overline{T}=1$.

Lazar et al \cite{Lazar} concluded that the exchange splitting of the
conduction band of the F-layer is the main physical reason for the strongly
reduced interface transparency. In principle, the calculation of the
interface transparency is a standard quantum mechanical problem of
reflection and transmission of electrons at the interface of two metals with
different Fermi energies. It is obvious that two electrons with opposite
spins forming a Cooper pair can never match the Fermi momenta of the
exchange-split subbands of a ferromagnet simultaneously and there will
always be a Fermi vector mismatch reducing the transmission. Additionally, a
decrease of $\overline{T}$ is expected due to a chemical mismatch of Pb and
Fe giving rise to a contact potential barrier at the interface. The barrier
height should be larger for immiscible metals like Pb and Fe in comparison
to metals that form solid solutions in the whole concentration range as,
e.g., V and Fe.

Measurements of dependence of $T_{c}$ on the thickness were
performed also for Fe/V/Fe trilayers \cite{Garifullin02}. For two
series of samples at small iron thicknesses the transition
temperature $T_{c}$ drops sharply when increasing $d_{Fe}$ up to
0.5 nm. Then, at $d_{Fe}\sim $0.7 nm for the series with
$d_{V}$=31 and 29 nm, a clear minimum of $T_{c}$ is observed. The
deepness of this minimum increases with decreasing $d_{V}$. For
these two series the residual resistivity ratio RRR$\simeq 4$,
meaning that the mean free path of the conduction electrons in the
S-layer $l_{S}\sim $ 4 nm \cite{Pippard}. The parameters resulting
from a theoretical fit of these curves are the superconducting
coherence length $\xi _{S}=\sqrt{\xi _{0}l_{S}/3.4}=4$ nm (here
$\xi _{0}$=44 nm is the BCS coherence length), as estimated from
the resistivity data. For the transparency parameter one obtains
$T_{m}=1.6$ and for the exchange length in the Fe film $\xi
_{h}=0.7$ nm.

Using the theoretical model calculations \cite{Tagirov+C} as a guideline,
one can extract the important physical parameters necessary for an
observation of the theoretically predicted rather spectacular re-entrant
behavior of the superconductivity, i.e. superconductivity vanishing for a
certain range of $d_{F}$ and coming back for larger $d_{F}$. The system
should possess a large electron mean free path in the F- as well as in the
S-layer, a high quantum-mechanical transparency of the S/F interface and a
geometrically flat interface without introducing too much diffuse scattering
of the electrons. The last two conditions are well fulfilled in Fe/V/Fe
trilayers, so one could try to further increase the electron mean free path $%
l_{F}$ or $l_{S}$. Whereas this is hardly possible for the F-layers since $%
l_{F}$ is limited by the very small layer thickness $d_{F}$, improved growth
conditions of the V-layer is a promising perspective to increase $l_{S}$
This was accomplished by samples prepared on single crystalline MgO (100)
with nearly epitaxial quality and an RRR-value of the order of 10. For this
set of samples clear re-entrant superconductivity was observed (Fig. 9) for
the first time.
\begin{figure}[th]
\centering{\includegraphics[width=6cm]{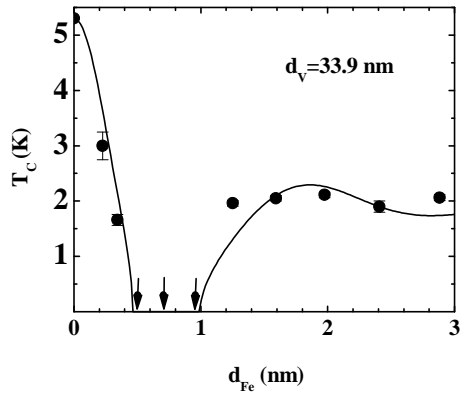}}
\caption{Superconducting transition temperature vs Fe thickness at fixed V
thickness for the Fe/V/Fe series with $d_{V}$=33.9 nm. The drawn line is a
theoretical curve with the parameters given in the text.}
\end{figure}
From the RRR-value for this series we estimate the mean free path of the
conduction electrons $l_{S}\sim 12$ nm and the corresponding coherence
length $\xi _{S}\sim 13$ nm. The latter was used for the theoretical fit of
the $T_{c}(d_{Fe})$-curve in Fig. 9. From the transparency parameter $%
T_{m}=1.6$ the average quantum mechanical transmission coefficient $%
\overline{T}$ \cite{Lazar} can be estimated to be ${\overline{T}}\simeq 0.6$%
. This value of $\overline{T}$ is about twice as large as the $\overline{T}$%
-value for the Pb/Fe interface discussed above \cite{Lazar}. This relatively
high transparency of the Fe/V interface is an essential ingredient for
observing re-entrance behavior. As mentioned above, a highly transparent S/F
interface is difficult to achieve with strong ferromagnets, since problems
with the matching of the Fermi momentum necessarily occur for at least one
spin direction.

An even higher transparency of the S/F interface can, in principle, be
achieved combining a superconductor with a ferromagnet weakened by dilution.
Recently \cite{Zdravkov}, the re-entrant superconductivity has been observed
for the Nb/Cu$_{1-x}$Ni$_{x}$ bilayers.

It is important to note that the results described above for the Fe/V system
were fitted using the values of the mean free path of the conduction
electrons in the F- and S-layers calculated from the resistivity data. In
contrast, for the Nb/Cu$_{1-x}$Ni$_{x}$ system Zdravkov et al \cite{Zdravkov}
had to use surprisingly large values for the mean free path for the
conduction electrons in the F-layer in their fitting procedure.

Finishing this section we would like to mention also study of the
interplay between magnetism and superconductivity in epitaxial
structures of half metal-colossal magnetoresistive
La$_{2/3}$Ca$_{1/3}$MnO$_{3}$ (LCMO) and
high-$T_{c}$ superconducting YBa$_{2}$Cu$_{3}$O$_{7-\delta }$ (YBCO) \cite%
{Jacob,Yeh,Sefrioui,Holden, Soltan}. Jacob et al \cite{Jacob}
demonstrated possibility of  preparation of hybrid perovskite
high-$T_{c}$ superconductor/ferromagnet superlattices. The
superlattices consisting of YBCO and LBMO
(La$_{2/3}$Ba$_{1/3}$MnO$_{3}$) layers with the thickness of a few
unit cells showed both strong colossal magnetoresistance at room
temperature and superconductivity at low temperatures.

Yeh et al \cite{Yeh} reported phenomena manifesting nonequilibrium
superconductivity induced by spin-polarized quasiparticles in
F/I/S (I is insulator) structures. Sefrioui et al \cite{Sefrioui},
basing on their measurements of $T_{c}$ vs S- and F-layer
thickness, speculate that injection of spin-polarized carriers
from LCMO into YBCO may add a new source of superconductivity
suppression: pair breaking by spin-polarized carriers. This pair
breaking effect extends over the spin diffusion length into S that
can be very long (it can be as long as 8 nm for YBCO). As a
result, in the YBCO layer superconductivity is suppressed by the
presence of manganite layers with a characteristic length scale
much longer than the one predicted by existing theories of the S/F
proximity effect.

The same result has been obtained by Holden et al \cite{Holden}
using ellipsometry measurements of the far-infrared dielectric
properties of superlattices composed of thin layers of YBCO and
LCMO. Finally, Soltan et al \cite{Soltan} studied the role of
spin-polarized self injection from LCMO into the YBCO layer. They
concluded that the nearly full spin polarization at the Fermi
level of LCMO leads to quenching of the proximity effect since it
prevents the Cooper pairs from tunnelling into the magnetic layer.
Thus, one can see that the results for superlattices consisting of
YBCO and LCMO presented above are somewhat contradicting to each
other. Nevertheless, they provide an avenue for future theoretical
studies of the F/S proximity effect in the presence of the
spin-polarized ferromagnets.

\subsection{ Superconducting spin valve}

In recent years much attention has been devoted to experimental
realization of the so-called superconducting spin valve. As
described in Sec. 2.1, a consequence of the S/F proximity effect
is that the superconducting transition temperature of a F/S/F
sandwich depends on the mutual orientation of the magnetization of
the two F-layers, the antiparallel orientation having a higher
$T_{c}$ than the parallel one \cite{Tagirov99}. In an ideal
superconducting spin valve the superconductivity of the S-layer
can be switched on and off by rotating the magnetization of one of
the F-layers relative to the other, giving an infinite
magnetoresistance for the switching field. The device is similar
to the well known conventional spin valve F/N/F system with a
normal metallic layer N interleaved between two ferromagnetic
layers F. In this device the antiparallel magnetization state
usually has a larger resistance than the parallel one.

It turned out that the realization of a superconducting spin valve
is difficult experimentally and the effects obtained until now are
quite small. There are two recent reports in the literature on the
successful realization a F/S/F superconducting spin valve. In the
CuNi/Nb/CuNi trilayer system \cite{Bader,Potenza} the maximum
shift by only 6 mK of the superconducting transition temperature
$T_{c}$ by changing the mutual orientation of the two
ferromagnetic layers from parallel to antiparallel was observed.
Actually, such a small shift may also be due to changes of the
domain structure of the ferromagnetic layers under the influence
of the external magnetic field \cite{Rusanov}.

Pena et al \cite{Pena} measured the magnetoresistance of F/S/F trilayers
combining the ferromagnetic manganite La$_{0.7}$Ca$_{0.3}$MnO$_{3}$ with the
high-$T_{c}$ superconductor YBa$_{2}$Cu$_{3}$O$_{7}$. They observed a
magnetoresistance in excess $1000\%$ for the superconducting state of YBa$%
_{2}$Cu$_{3}$O$_{7}$ that vanished in the normal state.

There is another possible design for the realization of the
superconducting spin valve effect proposed by Sungjun Oh et al
\cite{Sanjiun} that found less attention until now. It has the
layer structure S/F1/N/F2, i.e. two ferromagnetic layers F1 and F2
separated by a non-magnetic (N) layer are deposited on the one
side of the superconductor with F1 and N thin enough to allow the
superconducting pair wave function to penetrate into F2. The
authors have shown that changing the mutual magnetization
direction of F1
and F2 from parallel to antiparallel results in a substantial difference $%
\Delta T_{c}$ when the microscopic parameters for S- and F-films are
optimized.

For the realization of the F/S/F spin valve design
\cite{Tagirov99} it would be optimal to use a system where the
re-entrant $T_{c}(d_{F})$-behavior is observed. As discussed
above, Fe/V/Fe fulfills this criterion (see Fig. 9). However, an
acceptable performance of the spin valve with a sizable shift of
$T_{c}$ can only be expected if the S-layer thickness $d_{S}$\ is
close to the superconducting coherence length $\xi _{S}$. The
studies of the Fe/V/Fe system however revealed that the
superconductivity vanishes typically already at $d_{S}$\ $<3\xi
_{S}$. A possibility to overcome this problem and maintain
superconductivity at $d_{S}$\ $\sim \xi _{S}$ is to introduce very
thin non-ferromagnetic layers between the S- and F-layers that
should screen to some extent the very strong exchange field of the
F-layers.

A proper Fe/Cr/V/Cr/Fe system, where the Cr layers play the role of such
screening layers, has been studied in detail \cite{Garifullin04}. In Fig. 10
the $T_{c}$\ values measured for the samples from series with a fixed $%
d_{Fe} $=5 nm and $d_{Cr}$\ varied are plotted.
\begin{figure}[th]
\centering{\includegraphics[width=9cm]{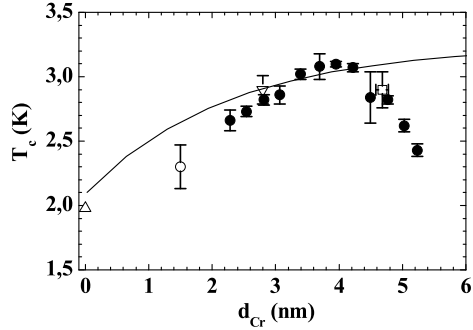}}
\caption{The superconducting transition temperature as a function of the
Cr-layer thickness for all samples from series (1). The solid line is a
theoretical curve (see main text).}
\end{figure}
In other three series $d_{Cr}$\ has been kept constant at $d_{Cr}$
=1.5, 2.8 and 4.7 nm and the thickness of the Fe-layer was varied.
The results for the transition temperatures of these series are
reproduced in Figs.
11b-11d and compared to previous results on Fe/V/Fe trilayers \cite%
{Garifullin02} (Fig. 11a).
\begin{figure}[th]
\centering{\includegraphics[width=7.6cm]{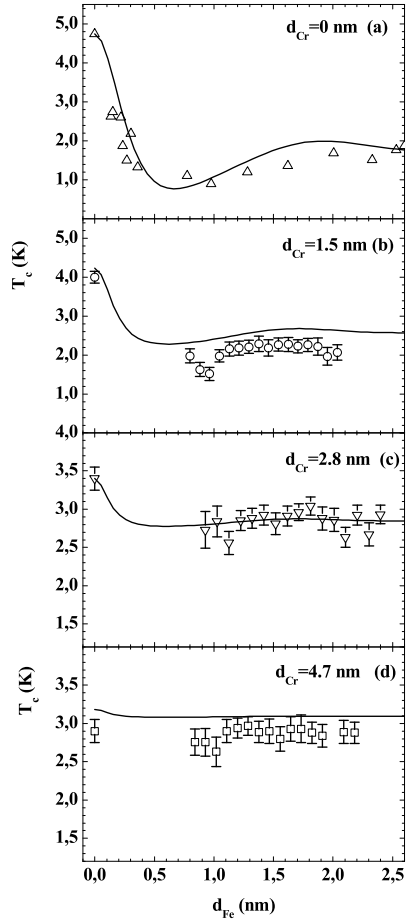}}
\caption{ Superconducting transition temperature as a function of the
Fe-layer thickness for samples from series with $d_{Cr}$\ =1.5 nm (b); with $%
d_{Cr}$\ =2.8 nm (c); with $d_{Cr}$\ =4.7 (d). The corresponding curve for
Fe/V/Fe trilayers is taken from Ref.~\protect\cite{Garifullin02} and shown
in (a) for comparison. The solid lines are calculations according to a model
explained in the main text.}
\end{figure}
The salient features of the results shown in Fig. 11 are as follows.

(1) The overall shape of the $T_{c}$\ ($d_{Fe}$\ )-curve is similar to that
obtained for Fe/V/Fe.

(2) The amplitude of the initial drop in $T_{c}$\ decreases with increasing
the thickness of the interleaved Cr-layer.

(3) At $d_{Cr}\ =4.7$ nm in Fig. 11d the Fe-layers have virtually
no influence on $T_{c}$ any more, indicating that the amplitude of
the pair wave function in the Fe-layer is negligible. This allows
the estimation of the penetration depth of the pair wave function
in Cr of about 4 nm, consistent with the results on Cr/V/Cr
trilayers \cite{Hubener}. These features are due to the expected
screening effect of the Cr-layer, since with increasing $d_{Cr}$\
the Cooper pair density reaching the Fe-layer is continuously
reduced and the effect of the strong exchange field in Fe on the
superconductivity is weakened.

The results of the model calculations are shown by the solid lines in Figs.
10 and 11a-11d and neglect the complications caused by the spin density wave
(SDW) state of antiferromagnetic Cr. The standard procedure described in the
literature (see, e. g., \cite{Hubener} and references therein) was applied
and the proximity effect of the V/Cr interface was treated by the
conventional theory for S/N metal films originally developed by de Gennes
\cite{deGennes}. In addition, pair breaking scattering of Abrikosov-Gor'kov
type \cite{AG} at magnetic defects in the Cr-layer is characterized by a
spin-flip scattering time $\tau _{s}$ i.e. Cr is treated as a paramagnetic
(P) layer.

Theory of the proximity effect for S/P/F layer system has been developed by
Vodopyanov et al \cite{Vodopyanov}. With certain assumptions \cite%
{Garifullin04} and the microscopic parameters known from the studies of the
Fe/V/Fe trilayers \cite{Garifullin02} all data points in Figs. 10 and 11
have been fitted simultaneously, with $\tau _{s}$ being the only fitting
parameter. All curves can be best described with $\tau _{s}=5\cdot 10^{-13}$
s. The overall shape of the curves is well reproduced, including the
penetration depth of about 4 nm for the superconducting pairing wave
function in Cr. This remarkably small penetration depth in Cr is thus
clearly proven to result from strong inelastic pair breaking scattering
leading to an exponential damping of the pair wave function amplitude within
the Cr-layer.

There is, however, an additional interesting experimental detail in Fig. 10
which the applied model fails to describe even in qualitative terms. This is
the drop of $T_{c}(d_{Cr})$ for $d_{Cr}$\ $\geq $4 nm, clearly seen in Fig.
10. This feature was attributed to a transition of the entire Cr-layer from
a non-magnetic state to an incommensurate SDW state at $d_{Cr}$\ $\sim $4
nm. The assumption of a strong suppression of the Cooper pair density by the
transition of the Cr layer from a non-magnetic to a spin density wave state
is plausible by the following reason. BCS-ordering and SDW-ordering in the
same region of the Fermi surface can be considered as competing electronic
ordering phenomena. In a theoretical paper on studying this problem (see,
e.g., \cite{Machida}), it was shown that those parts of the Fermi surface
where the nesting feature leads to a SDW state the formation of the BCS-gap
is suppressed and the superconducting transition temperature is reduced.

The study of the superconducting proximity effect in Fe/Cr/V/Cr/Fe gave new
results concerning the magnetic phase transition in the Cr-layer,
demonstrated a strong screening of the ferromagnetic exchange field of Fe by
the interleaved Cr-layers and allowed to estimate the upper limit of the
thickness of the screening Cr-layers for a spin valve to operate.

A novel approach for a realization of the superconducting spin valve design
originally proposed by Sungjun et al \cite{Sanjiun} was also undertaken
recently. The idea \cite{Westerholt} for the realization of such a device
was to choose as the non-magnetic interlayer N in the S/F1/N/F2/ layer
scheme an interlayer with a thickness corresponding exactly to an
antiferromagnetic interlayer exchange coupling between F1 and F2 \cite{Bruno}%
. Then, one can rotate the relative magnetization direction of F1 and F2
from antiparallel to parallel in an external field and observe the
accompanying shift of the $T_{c}$.

The experimental system of choice was the epitaxial superlattice
system MgO(100)/[Fe$_{2}$V$_{11}$]$_{20}$/V ($d_{V}$). (The index
denotes the number of monolayers.) There are several reasons that
make the choice of the epitaxial (V/Fe)-system favorable for
demonstrating the superconducting spin valve effect: First, it is
the superior quality of the Fe/V interface in the superlattice
\cite{Hjorvarsson,Andersson,Isberg,Uzdin} that guarantees a high
interface transparency and weak diffusive pair breaking scattering
at the interface. Second, the Fe$_{2}$ layers have a thickness
$d_{F}$ of about 0.3 nm only, whereas for the decay length of the
superconducting pair density $\xi _{F}\sim 0.7$ nm holds (see,
e.g., \cite{Lazar}). Thus the pair wave function within the
Fe$_{2}$-layer will only be weakly damped and the condition
$d_{F}/\xi _{F}<0.5$ optimal for observing the superconducting
spin valve effect will be fulfilled \cite{Sanjiun}.
\begin{figure}[th]
\centering{\includegraphics[width=7cm]{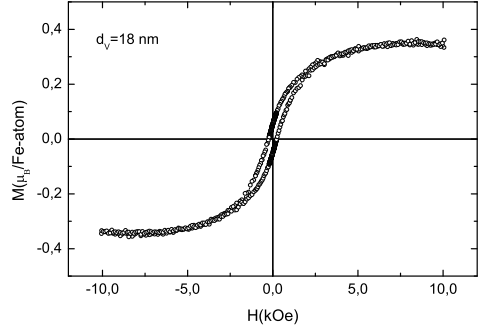}}
\caption{Magnetization hysteresis loop of the sample [Fe$_{2}$V$_{11}$]$%
_{20} $/V(18 nm) measured at 10 K. }
\end{figure}
In Fig. 12 we reproduce the magnetization curve of a [Fe$_{2}$V$_{11}$]$%
_{20} $- superlattice measured at 10K. The shape of the hysteresis
shows that the interlayer exchange coupling is antiferromagnetic
with a ferromagnetic saturation field of $H_{sat}=6$ kOe. The
upper critical magnetic field for the field direction parallel and
perpendicular to the film plane is plotted in Fig. 13 for several
samples.
\begin{figure}[th]
\centering{\includegraphics[width=6.5cm]{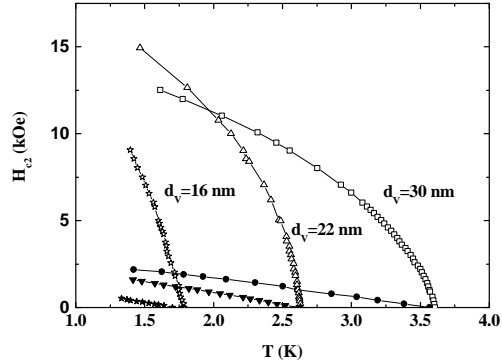}}
\caption{Upper critical magnetic field versus temperature with the field
applied parallel and perpendicular to the film plane for three samples [Fe$%
_{2}$V$_{11}$]$_{20}$/V($d_{V}$). The thickness $d_{V}$ is given in the
figure, the open symbols refer to the magnetic field direction parallel to
the film plane, the solid symbols refer to the direction perpendicular to
the plane. }
\end{figure}
For a two dimensional (2D) thin film with the magnetic field perpendicular
or parallel to the film plane the classic result for the upper critical
field is \cite{Tinkham}:
\begin{equation}
H_{c2}^{perp}={\frac{\Phi _{0}}{{2\pi \xi ^{2}(0)}}}{\left( {1-{\frac{T}{{%
T_{c}}}}}\right) }
\end{equation}%
\begin{equation}
H_{c2}^{par}={\frac{\Phi _{0}}{{2\pi \xi (0)}}}{\frac{\sqrt{12}}{{d_{s}}}}%
\sqrt{\left( {1-{\frac{T}{{T_{c}}}}}\right) }
\end{equation}%
with the flux quantum $\Phi _{0}$, the thickness of the film $d_{S}$\ and
the Ginzburg-Landau correlation length $\xi (0)$ related to Pippard's
correlation length $\xi _{s}$ as $\xi (0)=1.6\xi _{s}$.

The measurements of the upper critical field for Fe/V/Fe trilayers for
parallel orientation of the magnetic field relative to the film plane is
perfectly described by Eq. (16), as it was observed earlier \cite%
{Wong,Koorevaar}.
\begin{figure}[th]
\centering{\includegraphics[width=7cm]{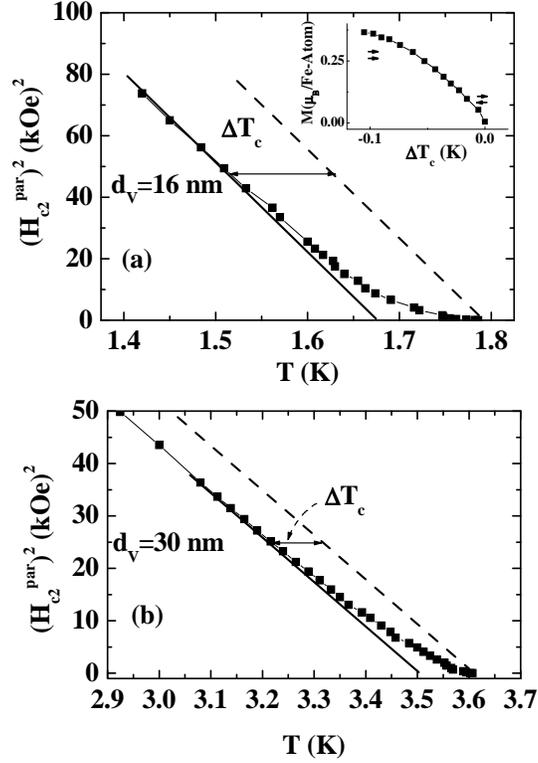}}
\caption{Square of the parallel upper critical magnetic field versus
temperature for the sample [Fe$_{2}$/V$_{11}$]$_{20}$/V(16 nm) (a) and [Fe$%
_{2}$/V$_{11}$]$_{20}$/V(30 nm) (b). The full straight line is the
extrapolation of the linear temperature dependence for higher fields, the
dashed line is the theoretical curve expected if the magnetization of the
superlattice would not change. $\Delta T_{c}$ is the shift of the
superconducting transition temperature between the superlattice in the
antiferromagnetic state and in ferromagnetic saturation. The inset in panel
(a) depicts the shift of the superconducting transition temperature with the
magnetization of the [Fe$_{2}$/V$_{11}$]$_{20}$ superlattice}
\end{figure}
In Figs. 14a and 14b the square of parallel upper critical field
are plotted together with the straight line that describes the
temperature dependence for fields above 6 kOe perfectly. Below
$H=6$ kOe there is an increasing deviation from the straight line.
From the extrapolation of the straight line one gets a
superconducting transition temperature $T_{c}$\ that is more than
0.1K below the true transition temperature measured at zero field.
A comparison with the magnetization curve in Fig. 12 shows that
the ferromagnetic saturation field of 6 kOe is correlated with the
first deviation of $H_{c2}^{2}(T)$ from the straight line in Fig.
14a and 14b. From this one can infer that the deviation of the
upper critical field from the 2D-behavior in Fig. 14 is caused by
the gradual change of the sublattice magnetization direction of
the [Fe$_{2}$/V$_{11}$]$_{20}$-superlattice from
parallel above 6 kOe to antiparallel in zero field. For the sample with $%
d_{V}$\ =16 nm in Fig. 14a $T_{c}$ =1.78K in the antiferromagnetic
state, while in the ferromagnetic saturation we extrapolate
$T_{c}$=1.67K. The temperature difference $\Delta T_{c}$=0.11K is
the anticipated superconducting spin valve effect.

These experiments clearly demonstrate that the superconducting transition
temperature of the V-film reacts sensitively on the mutual magnetization
orientation of the Fe$_{2}$ layers of an antiferromagnetically coupled [Fe$%
_{2}$V$_{11}$]$_{20}$ superlattice. Actually the ferromagnetic layers in
this system cannot be switched from the parallel to the antiparallel state,
since the parallel state needs the application of a strong external magnetic
field. At the same time, it should be in principle possible to construct a
switching device by replacing the antiferromagnetically coupled [Fe/V]
superlattice by a conventional spin valve trilayer system.

\subsection{Odd triplet superconductivity in S/F/S structures}

In section 2.3 a theoretical model predicting a possible robust
triplet proximity effect in S/F structures has been described. The
mechanism is operational in the presence of a rotating
magnetization at the S/F interface. Recently Sosnin et al
\cite{Sosnin} presented the first clear experimental indication of
this type of proximity effect using an Andreev type of
reflectometer and an S/F/S mesoscopic thin film structure. The
design of the reflectometer is depicted schematically in Fig. 15a.
\begin{figure}[th]
\centering{\includegraphics[width=9cm]{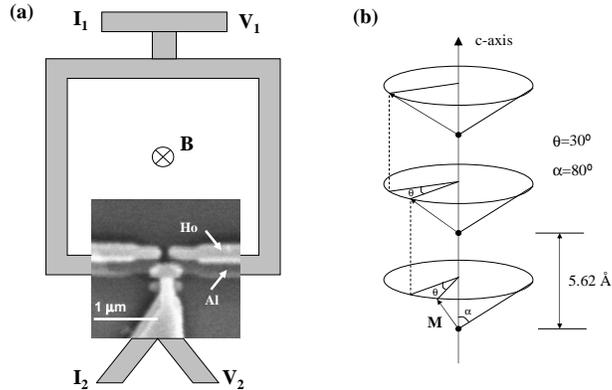}}
\caption{a) Experimental set-up and SEM micrograph of S/F/S junction area
prepared by shadow evaporation. b) Magnetic structure of Ho: magnetization
\textbf{M} rotates by 30$^{o}$ each atomic layer along \textit{c}-axis at an
angle of 80$^{o}$ to this axis.}
\end{figure}
It consists of a superconducting Al-loop with an area of 20
$\mu$m$^{2}$ with a narrow gap bridged by a ferromagnetic
Ho-stripe. The distance between the two Al/Ho contact points was
more than one order of magnitude larger than the singlet magnetic
coherence length $\xi _{F0}$. A rotating magnetization at the S/F
interface is established here by the intrinsic conical
ferromagnetism of Ho (see Fig. 15 b). The essential experimental
finding is that below the critical temperature $T_{c}$ of Al the
resistance of the Ho wire exhibits oscillations as a function of
the superconducting phase difference between the two interfaces of
the Ho-stripe with the superconducting Al ring, as shown in Fig.
16a. The phase difference was generated by varying the magnetic
flux penetrating the Al-loop.
\begin{figure}[th]
\centering{\includegraphics[width=6cm]{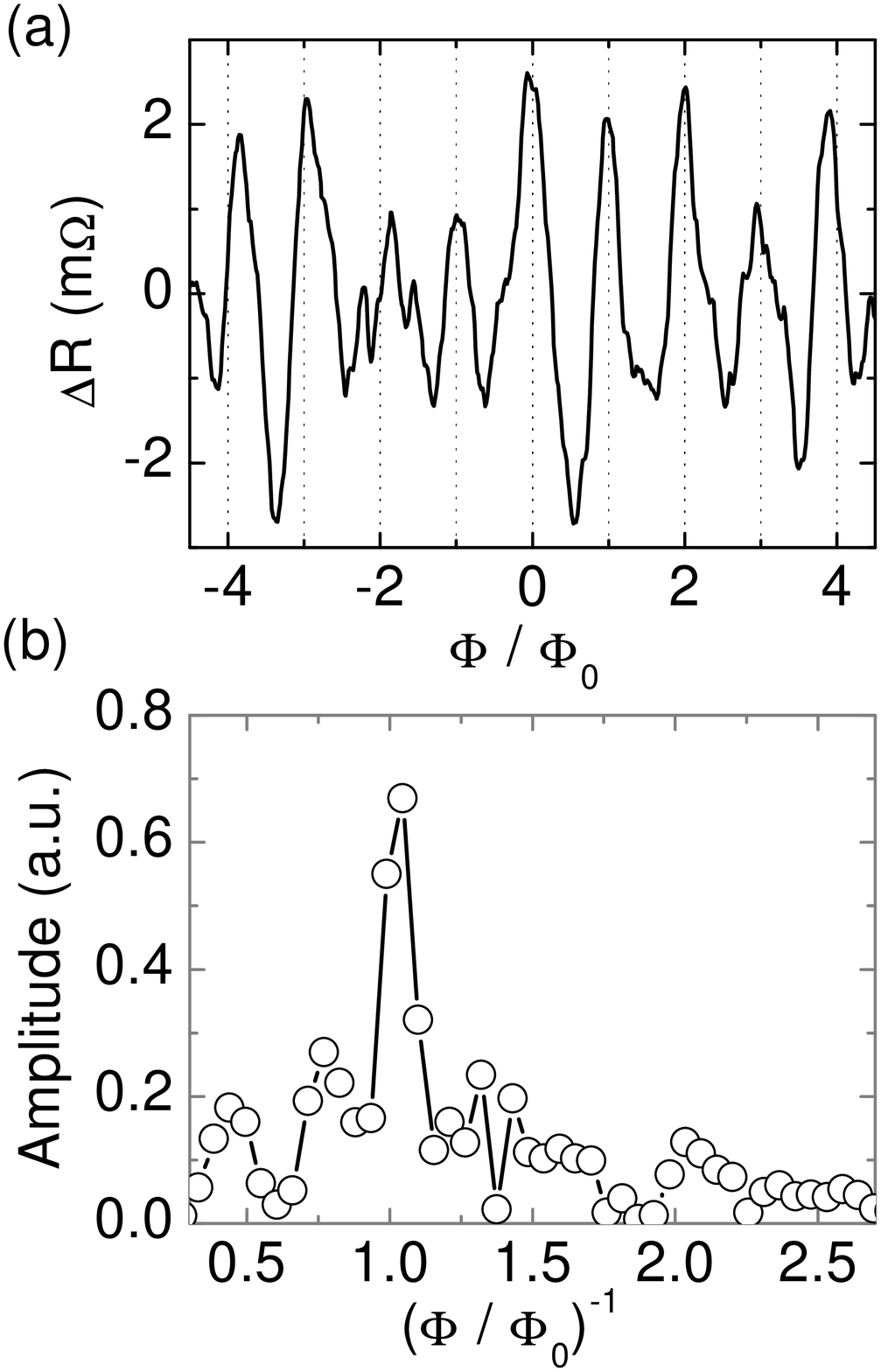}} \caption{a)
Magnetoresistance oscillations of the sample shown in Fig. 15a
measured at T=0.27 K as a function of normalized external flux
through the loop. The sample resistance is 94.3 $\Omega $. b)
Fourier spectrum of the oscillations confirming the $hc/2e$
periodicity.}
\end{figure}
The period of the oscillations corresponds to the flux quantum $\Phi
_{0}=2\cdot 10^{-7}$ Gcm$^{2}$ and gives rise to the sharp peaks in the
Fourier spectrum of oscillations (Fig. 16b). Estimates show that for the
relative amplitude of the conductance oscillations $\Delta R/R_{F}\simeq
10^{-4}$ ( $R_{F}$ is the resistivity of the ferromagnetic wires) is
expected. These oscillations were observed for the samples with a distance
between the Al/Ho contact points interfaces of up to $L_{F}=160$ nm. Such a
long-range phase coherence cannot be explained by the proximity effect
involving the penetration of the ordinary singlet pairs, since the upper
limit for the singlet penetration depth $\xi _{F0}$ is equal to the electron
mean free path $l$ which for was $l\approx 6$ nm. Thus, the observed
oscillations of the magnetoresistance seem to originate from the long-range
penetration of a helical triplet component of superconductivity generated in
a ferromagnetic conductor and induced by the presence of a rotating
magnetization.

Recently Keizer et al \cite{Keizer} studied lateral S/F/S
Josephson junctions combining the strong ferromagnet CrO$_{2}$
that belongs to the group of half-metals with full spin
polarization of the electrons at the Fermi level and the
conventional s-wave superconductor NbTi. They observed a Josephson
supercurrent prevailing over very long length scales up to $\sim $
1 $\mu$ m. This is by orders of magnitude larger than expected for
singlet correlations, which is of the order of 1 nm. In addition
to the long-range penetration of the superconducting pair density
into CrO$_{2}$, they found that the supercurrent strongly depended
on the magnetization direction in the ferromagnet. On the basis of
these findings Keizer et al attributed the long-range supercurrent
to the triplet correlations. In this case of a half-metallic
ferromagnet it is reasonable to assume that the LRTC is created at
the S/F interface where spin-flip processes may happen
\cite{Schoen1}.

Hints on the realization of the triplet proximity effect also came from
recent magnetization data on hybrid structures consisting of multilayers of
manganites [La$_{0.33}$Ca$_{0.67}$MnO$_3$/La$_{0.60}$Ca$_{0.40i}$MnO$_3$]$%
_{15}$ in contact with a low-$T_c$ Nb superconductor \cite{Stamopoulos}.

\subsection{Other proximity effects}

It seems natural and actually it is theoretically well established that the
penetration of superconductivity from the S- into the F-layers is not the
only possible proximity effect in S/F systems (see section 2.5, 2.6). The
proximity effect can also work in the reverse direction, i.e. the
ferromagnetism from the F-layer can leak into the S-layer (inverse proximity
effect) or the S-layer can modify the ferromagnetic state of the F-layer
(cryptoferromagnetism). However, these effects are more subtle from the
experimental point of view and are sill less established.

\subsubsection{Cryptoferromagnetism in S/F layers}

As shown in Sec. 2.5, under certain conditions the ferromagnetic order in
F-layers may be reconstructed by the action of the S-layer into a new
magnetic domain state \cite{Buzdin88} or a cryptoferromagnetic state \cite%
{BEL}. The basic physical reason for this behavior is that the destructive
influence of the ferromagnetic exchange field on the superconductivity can
be considerably reduced if the ferromagnetic state is modified in such a
manner that the exchange field cancels when averaged over the
superconducting coherence length.

The first hint in favor for a reconstruction of the ferromagnetic
state below the superconducting transition temperature was
obtained from the anomalous temperature dependence of the
ferromagnetic resonance (FMR) line width observed in epitaxial
Fe/Nb bilayers below $T_{c}$ \cite{Muhge98}. However,
a quantitative estimate using the theory of Buzdin and Bulaevskii \cite%
{Buzdin88} rase doubts in this interpretation, since the effect in Fe/Nb
should only occur at an Fe-layer thickness an order of magnitude smaller
than observed experimentally. Later Bergeret et al \cite{BEL} studied
theoretically the possibility of a non-homogeneous magnetic order of a
ferromagnetic film placed on top of a bulk superconductor. They also
concluded that due to the large magnetic stiffness constant in Fe, the
cryptoferromagnetic state can hardly be realized using pure Fe films. These
considerations suggested that the tendency to a reconstruction of the
ferromagnetic state observed experimentally in Fe/Nb might be caused by a
granular structure of the very thin Fe layers.

Quantitative estimates by Bergeret et al \cite{BEL} showed that the
transition from the ferromagnetic to the cryptoferromagnetic state should be
observable in a ferromagnet with a magnetic stiffness constant an order of
magnitude smaller than that of pure Fe. This can be achieved by dilution of
Fe in suitable alloy systems, a favorable choice being Pd$_{1-x}$Fe$_{x}$ at
small x due to its low and tunable Curie temperature.

In an FMR study for a series of samples V/Pd$_{1-x}$Fe$_{x}$ the
temperature
dependence of the effective magnetization $4\pi M_{eff}=4\pi M-(2K_{u}/M)$ ($%
M$ is the saturation moment of the ferromagnet and $K_{u}$ is the
perpendicular anisotropy constant) was measured \cite{Garifullin2}. The
low-temperature part of $4\pi M_{eff}(T)$ is depicted in Fig. 17.
\begin{figure}[th]
\centering{\includegraphics[width=6cm]{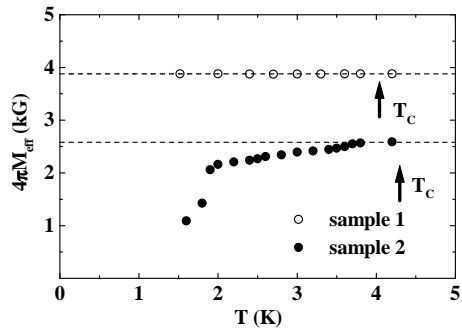}}
\caption{Low-temperature parts of $4\protect\pi M_{eff}(T)$ for the sample 1
with $d_{V}$=37.2 nm, $d_{Pd-Fe}$=4.4 nm, $T_{c}$\ =4.0 K and sample 2 with $%
d_{V}$=40 nm, $d_{Pd-Fe}$=1.2 nm, $T_{c}$\ =4.2 K. The arrows show the $%
T_{c} $-values at the resonance field $H_{0}$.}
\end{figure}
One observes a decrease of the effective magnetization $4\pi M_{eff}$ below $%
T_{c}$ for the sample 2 (Fig. 17) but not for sample 1. A decrease of $4\pi
M_{eff}$ can be caused by a decrease of the saturation magnetization $M$ or
by an increase of the perpendicular uniaxial anisotropy constant $K_{u}$. A
comparison of FMR results for films with different thickness of the
ferromagnetic layers leads to the conclusion that $K_{u}$ is very small and
the decrease of $4\pi M_{eff}$ must be caused by a decrease of the
saturation magnetization $M$. This suggests that the decrease of the
saturation magnetization below $T_{c}$ is caused by a reconstruction of the
ferromagnetic state.

An estimate following the phase diagram by Bergeret et al
\cite{BEL} (Fig. 5) gave the parameters $a\sim 1.2$ and $\lambda
\sim 1.3\cdot 10^{-3}$ for sample 2 with $d_M\sim$1.2 nm and
$T_{Curie}\sim$100K. In accordance with the phase diagram of
Bergeret et al (Fig. 5) this implies that starting from $\tau \sim
0.2$ ($T\sim 3.2$ K) a transition from the ferromagnetic to the
cryptoferromagnetic state should take place,
as it is actually observed experimentally. For the sample $1$ with $%
d_{M}\sim$ 4.4 nm and $T_{Curie}\sim $ 250 K we have $a\sim 20$
and $\lambda \sim 1.4\cdot 10^{-2}$. With these parameter values
the ferromagnetic state should be stable at any temperature, in
agreement with the experimental result.

Thus, these estimates support the conclusion that a phase transition from
the ferromagnetic state to the cryptoferromagnetic state occurs in sample 2.
However, one cannot completely exclude that the anomalous temperature
dependence of $M_{eff}$ might be due to the screening of the magnetic
moments of the ferromagnetic layer by the polarized Cooper pairs, as
discussed in Sec. 2.6 and in the next section.

\subsubsection{Inverse proximity effect}

Up to now any unequivocal experimental evidence for the penetration of the
magnetization from the ferromagnetic side into the superconducting side of
an S/F bilayer, as discussed in Sec. 2.6, does not exist. First
interpretations in this direction have been published only recently \cite%
{Stamopoulos,Stahn}. Stahn et al \cite{Stahn} studied the
magnetization profile of
[YBa$_{2}$Cu$_{3}$0$_{7}$/La$_{2/3}$Ca$_{1/3}$MnO$_{3}$]
multilayers using neutron reflectometry. From a change of the
reflectivity curves below the superconducting transition
temperature they could propose two possible magnetization
profiles. In the first one a magnetic moment within the
superconducting layer antiparallel to that of the F-layer exists,
which is consistent with the inverse proximity effect. The second
profile had a magnetically "dead" interlayer region in the F-layer
with zero magnetic moment. Stahn et al \cite{Stahn} argue in favor
of the first possibility but the situation is not yet settled.

Stamopoulos et al \cite{Stamopoulos} presented magnetization measurements on
multilayers of manganites [La$_{0.33}$Ca$_{0.67}$MnO$_{3}$/La$_{0.60}$Ca$%
_{0.40}$MnO$_{3}$]$_{15}$ in contact with a low-$T_{c}$
superconductor. They came to the conclusion that the
superconductor below $T_{c}$ becomes ferromagnetically coupled to
the multilayer. Since it is expected that for the singlet pairing
the magnetization of F penetrates into S antiferromagnetically,
the authors conclude that a spin-triplet superconducting component
forms and penetrates into the F-layer thus inducing the
ferromagnetic coupling observed experimentally.

\section{Summary and Conclusions}

The main purpose of the present paper was to review the status of the
research on the proximity effects in the S/F  layer systems from the
experimental as well as from the theoretical point of view.

Peculiarities of the S/F proximity effect originating from the penetration
of the condensate function into the ferromagnet that have been discussed
controversially in the beginning seem to be well established by now. The
S/F/S Josephson junctions with the $\pi $-coupling are, e.g., even suggested
as basic units for realization of Q-bits for quantum computing \cite{Blatter}%
.

It has become already traditional in the field of the S/F
proximity effect that theory is somewhat ahead of experiment. The
situation persists and intriguing theoretical predictions are
still waiting for the first experimental verifications or further
experimental support. One of these predictions concerns the
unconventional superconductivity in S/F systems. The experimental
realization is difficult, since the unconventional
superconductivity expected here, namely odd triplet
superconductivity, can only be generated by a rotating
magnetization at the interface. Nevertheless, the first
experimental indications of its existence have already been
reported. What is important, the odd triplet superconductivity is
insensitive to scattering on non-magnetic impurities and this is
certainly helpful for an experimental observation.

We should also mention further recent ideas on how to identify the
odd-triplet superconductivity \cite{tanaka1,tanaka2,nazarov}.

The inverse proximity effect, i.e. the penetration of the magnetic order
parameter into a superconductor, has not been clearly observed
experimentally until now. However, indications on the closely related
effect, namely the decreasing of the total ferromagnetic moment below $T_{c}$%
, already exist. Yet, it is not easy to clarify to what extent the
non-homogeneous distribution of $M_{F}$ produced in the ferromagnet below $%
T_{c}$ contributes to the effect. The best way to observe the spin screening
of $M_{F}$ is either probing directly the spatial distribution of the
magnetic field using neutron scattering or by measuring muon spin resonance.
Since the magnetic moment $M$ varies on the macroscopic length $\xi _{S}$,
it should be possible to detect it.

A considerable work still remains to be done on the experimental side. Only
careful material selection, optimization of film preparation and device
design will enable one clarification of all the complex phenomena that may
occur in the S/F proximity systems. Very promising seems the study of F/S
structures with comparable ferromagnetic Curie and superconducting
transition temperatures. Combining elemental superconductors with elemental
ferromagnets, as was done in the majority of papers on the S/F proximity
effect published until now, is not the best way for the observation of the
proximity effects because the ferromagnetic exchange energy is orders of
magnitude larger than the superconducting condensation energy. In this case
the ferromagnetic state can hardly be modified by the superconductor. Rare
earth based ferromagnetic compounds with low Curie temperatures would, in
principle, be a better choice. Combining high-$T_{c}$ superconductors with
ferromagnetic oxides is another promising option. \newline

\textit{The authors are grateful for the support by the Deutsche
Forschungsgemeinschaft (DFG) within SFB 491. One of us (AFV) would like to
acknowledge financial support from DFG within the project
Mercator-Gastprofessoren.}




\end{document}